\documentclass[onecolumn,showpacs,preprintnumbers,showkeys]{revtex4}
\usepackage{amsthm, amscd, amsfonts, amssymb, graphicx, hhline}
\usepackage{dcolumn}
\usepackage{bm}
\usepackage[english]{babel}
\usepackage{footnpag}
\usepackage{nccfoots}
\usepackage{multirow}
\usepackage[symbol*]{footmisc}

\begin{document}

\title{The role of electron-vibron interaction and local pairing
in conductivity and superconductivity of alkali-doped fullerides
}
\author{Konstantin V. Grigorishin}

\email{konst.phys@gmail.com} \affiliation{Boholyubov Institute for
Theoretical Physics of the National Academy of Sciences of
Ukraine, 14-b Metrolohichna str. Kiev-03680, Ukraine.}

\begin{abstract}

We investigate the competition between the electron-vibron
interaction (interaction with the Jahn-Teller phonons) and the
Coulomb repulsion in a system with the local pairing of electrons
on the 3-fold degenerate lowest unoccupied molecular orbital
(LUMO). The el.-vib. interaction and the local pairing radically
change conductivity and magnetic properties of alkali-doped
fullerides $\texttt{A}_{n}\texttt{C}_{60}$, which would have to be
antiferromagnetic Mott insulators: we have shown that materials
with $n=1,5$ and $\texttt{A}=\texttt{K},\texttt{Rb}$ are
conductors but not superconductors; $n=3$ and
$\texttt{A}=\texttt{K},\texttt{Rb}$ are conductors and
superconductors at low temperatures, but with
$\texttt{A}=\texttt{Cs}$ they are Mott-Jahn-Teller insulators at
normal pressure; $n=2,4$ are nonmagnetic Mott insulators. Thus
superconductivity, conductivity and insulation of these materials
have common nature. Using this approach we obtain the phase
diagram of $\texttt{A}_{3}\texttt{C}_{60}$ analytically, which is
the result of interplay between the local pairing, the el.-vib.
interaction, Coulomb correlations, and formation of small radius
polarons.
\end{abstract}


 \maketitle

\section{Introduction}\label{intr}

\begin{figure}[h]
\includegraphics[width=10cm]{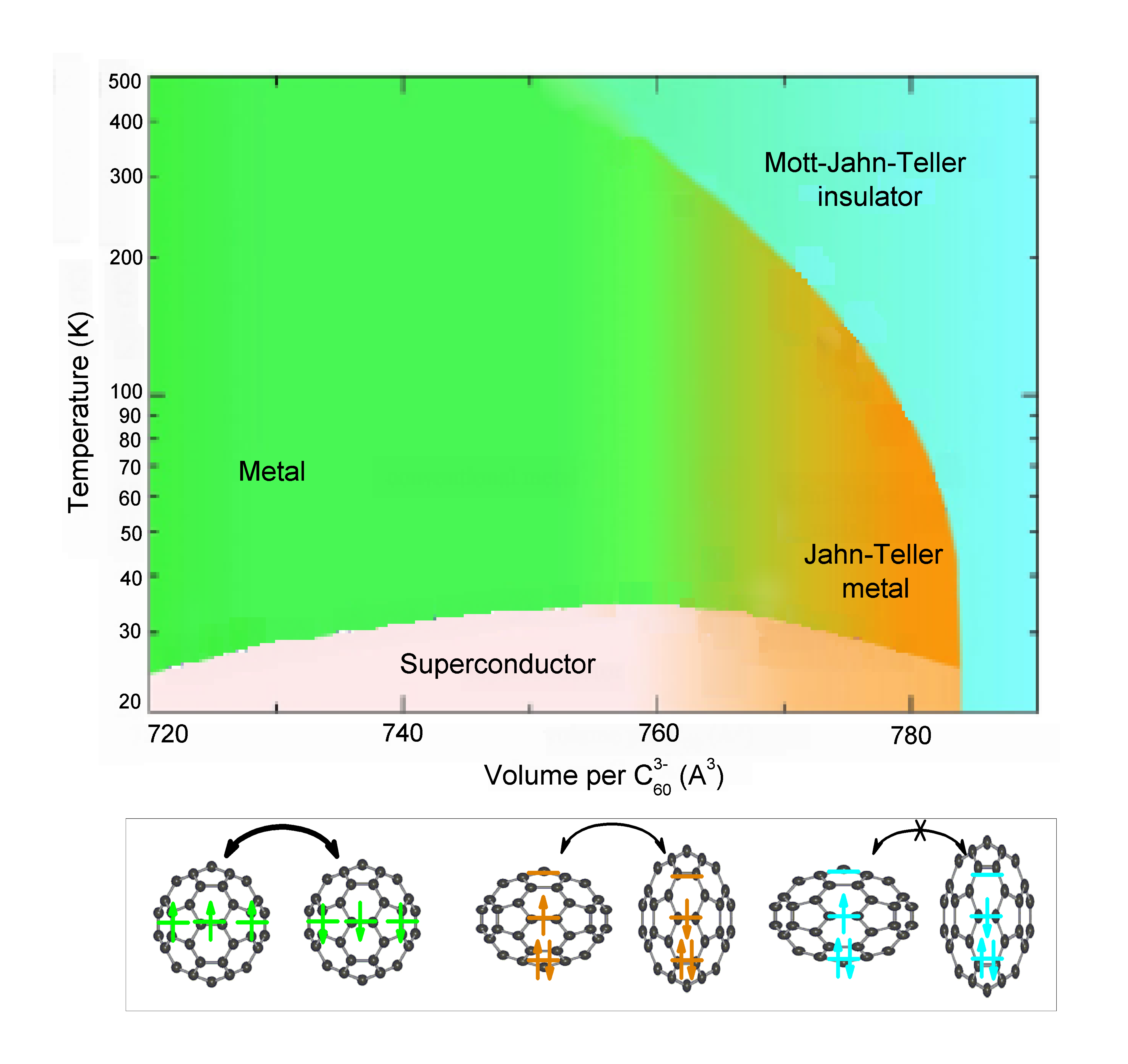}
\caption{Experimental phase diagram of fcc-structured
$\texttt{Rb}_{x}\texttt{Cs}_{3-x}\texttt{C}_{60}$, as a function
of volume per $\texttt{C}_{60}$ experimentally obtained in
\cite{zadik}. Within the metallic (superconducting) regime,
gradient shading from orange to green schematically illustrates
crossover from the JTM to conventional metal. Lower panel:
schematic depictions of the respective molecular electronic
structure, intermolecular electron hopping, and JT molecular
distortion for conventional metal, JT metal and MJT insulator in
their interpretation.} \label{Fig1}
\end{figure}

Alkali-doped fullerides ($\texttt{A}_{n}\texttt{C}_{60}$ with
$\texttt{A}=\texttt{K},\texttt{Rb},\texttt{Cs}$ and $n=1\ldots 5$)
demonstrate surprising properties. Simple band theory arguments
predict that any partial filling between 0 and 6 electrons
(between empty and full molecular orbital $t_{1u}$, respectively)
should give a metallic behavior. In the same time these materials
are characterized with a narrow conduction band $W\sim 0.3\ldots
0.5\texttt{eV}$ and a strong on-site Coulomb repulsion $U\sim
0.8\ldots 1.0\texttt{eV}$. Moreover electrons on $t_{1u}$
molecular orbital should be distributed according to Hund's rule:
spin of a molecule must be maximal. Thus alkali-doped fullerides
should be antiferromagnetic Mott insulators (MI). In reality the
properties of the alkali-doped fullerides are in striking
contradiction to the expected they. So
$\texttt{A}_{2}\texttt{C}_{60}$ and
$\texttt{A}_{4}\texttt{C}_{60}$ are nonmagnetic insulators. Thus
molecule $\texttt{C}_{60}$ with additional electrons in LUMO does
not have spin, that contradicts to Hund's rule. Under pressure
these materials become metallic. $\texttt{A}_{1}\texttt{C}_{60}$
and $\texttt{A}_{5}\texttt{C}_{60}$ are conductors.
$\texttt{A}_{3}\texttt{C}_{60}$ are superconductors with
$\texttt{A}=\texttt{K},\texttt{Rb}$ for which the critical
temperatures are sufficiently high $T_{c}\sim 30\texttt{K}$.
However for $\texttt{A}=\texttt{Cs}$ the material is insulator,
but it becomes superconductor under high pressure $\sim
2\texttt{kbar}$. The corresponding phase diagram of
$\texttt{A}_{3}\texttt{C}_{60}$ is shown in Fig.(\ref{Fig1}).
These materials at low temperature are superconductors with
dome-shaped $T_{c}$ versus lattice constant, and they are
conductors for higher temperature. However at large lattice
spacing and at hight temperature these phases are broken off with
a Mott insulating phase. The insulating phase is magnetic
Mott-Jahn-Teller (MJT) insulator (antiferromagnetic with
$T_{N}=46\texttt{K}$ for the $\texttt{A15}$ structure and
$T_{N}=2.2\texttt{K}$ for the $\texttt{fcc}$ structure or with a
spin freezing only below $10\texttt{K}$ due to frustration of the
$\texttt{fcc}$ lattice), with the on-molecule distortion creating
the ground state with spin $S=1/2$, which produces the magnetism.
Results of infrared spectroscopy \cite{zadik,klupp,kamar,bald} are
interpreted \cite{zadik,takab} as that the insulator-to-metal
transition is not immediately accompanied by the suppression of
the molecular Jahn-Teller distortions. The metallic state that
emerges following the destruction of the Mott insulator is
unconventional - sufficiently slow carrier hopping and the
intramolecular Jahn-Teller (JT) effect coexists with metallicity.
This JT metallic state of matter demonstrates both molecular
(dynamically JT-distorted $\texttt{C}_{60}^{3-}$-ions observed in
\cite{klupp,kamar}) and free-carrier (electronic continuum)
features. As the fulleride lattice contracts further, there is a
\emph{crossover} from the JT metal to a conventional Fermi liquid
state upon moving from the Mott boundary towards the
under-expanded regime, where the molecular distortion arising from
the JT effect disappear and the electron mean free path extends to
more than a few intermolecular distances.  However, it should be
noted, since the line of phase transition from JT metal to
conventional metal is absent and the molecular distortions
\emph{gradually} increase as lattice expands, that the JT metal is
not a phase, unlike the MJT insulator which is separated from the
metal state by a line of the first kind phase transition.
Spin-lattice relaxation measurements \cite{alloul1,alloul2} show
that insulators $\texttt{Na}_{2}\texttt{C}_{60}$ and
$\texttt{K}_{4}\texttt{C}_{60}$ have a nonmagnetic ground state
and their low energy electronic excitations are characterized by a
spin gap (between singlet and triplet states of the Jahn-Teller
distorted $\texttt{C}_{60}^{2-}$ or $\texttt{C}_{60}^{4-}$),
furthermore, it has been evidenced very similar electronic
excitations in $\texttt{Na}_{2}\texttt{Cs}\texttt{C}_{60}$ and
$\texttt{Rb}_{3}\texttt{C}_{60}$, coexisting with typically
metallic behavior: the $\texttt{C}_{60}^{2-}$ and
$\texttt{C}_{60}^{4-}$ would be formed within the metal on very
short time scales $(10^{-14} s)$ that do not imply static charge
segregation. These facts speak  about a tendency to charge
localization in the odd electron system and the dynamic JT
distortions can induce attractive electronic interaction in these
systems. $\texttt{Cs}_{3}\texttt{C}_{60}$ exists in two
polymorphs: an ordered $\texttt{A}15$ structure and a merohedrally
disordered $\texttt{fcc}$ one similar to
$\texttt{K}_{3}\texttt{C}_{60}$ and
$\texttt{Rb}_{3}\texttt{C}_{60}$. Though the ground state
magnetism of the Mott phases differs, their high $T$ paramagnetic
and superconducting properties are similar, and the phase diagrams
versus unit volume per $\texttt{C}_{60}$ are superimposed
\cite{alloul3,alloul4,alloul5}. Due to the crystal field,
determined by the potential created by the neighboring
$\texttt{Cs}^{+}$ ions,  the energies of molecular distortions,
that are equivalent in a free molecular ion, can differ in a
crystal when pointed at different crystallographic directions. The
crystal field influences the vibrational levels of
$\texttt{C}_{60}^{3-}$ molecular ions making the spectra
temperature- and polymorph-dependent, the presence of
temperature-dependent solid-state conformers validates the proof
of the dynamic Jahn–Teller effect \cite{kamar}. Thus the crystal
field can additionally induce and stabilize the JT distortions
that much stronger manifests in the $\texttt{A}15$ structure
compared with the disordered $\texttt{fcc}$ structure.

The mechanism of superconductivity of the alkali-doped fullerides
has not been fully understood. The positive correlation between
$T_{c}$ and the lattice constant found in $\texttt{K}$- and
$\texttt{Rb}$-doped fullerides has been understood in terms of the
standard BCS theory: the density of states of conduction electrons
is $\nu\propto 1/W$, and
$T_{c}=1.14\langle\omega\rangle\exp(-1/\lambda\nu)$, thus the
smaller $W$ the large $T_{c}$. Therefore superconductivity of
$\texttt{A}_{3}\texttt{C}_{60}$ is often described with Eliashberg
theory in terms of electron-phonon coupling and Tolmachev's
pseudopotential $\mu^{\ast}$ \cite{gun0,chen1,cap}. In the same
time there is another approach to describe phases of alkali-doped
fulleride - the model of \emph{local pairing}
\cite{han1,han2,lam1,lam2,koga}. The experimental basis for this
hypothesis is the fact that the coherence length (size of a Cooper
pair) in the superconducting alkali-doped fulleride is $\sim
2\ldots 3\texttt{nm}$, which is comparable with a size of a
fullerene molecule $\texttt{C}_{60}$ $\sim 1\texttt{nm}$.
Moreover, the Hubbard-like models predict that
$\texttt{A}_{4}\texttt{C}_{60}$ is an anti-ferromagnetic
insulator, while it is known experimentally that there are no
moments in $\texttt{A}_{4}\texttt{C}_{60}$. The electrons in
$\texttt{A}_{n}\texttt{C}_{60}$ relatively strong interact with
Jahn-Teller intramolecular phonons with $H_{g}$ and $A_{g}$
symmetries. The interaction favors a low-spin state and might lead
to a nonmagnetic insulator. At the same time there is, however, a
Hund's rule coupling, which favors a high-spin state. Thus
competition between the Jahn-Teller coupling and the Hund's rule
coupling takes place. As a result each molecule
$\texttt{C}_{60}^{3-}$ in $\texttt{Cs}_{3}\texttt{C}_{60}$ crystal
(antiferromagnetic insulator) has spin $S=1/2$ instead $3/2$.
Proceeding from these facts the local pairing model suggests that
the electron-vibron interaction (interaction of electrons with
$H_{g}$ and $A_{g}$ intramolecular Jahn-Teller oscillations)
favors the formation of a local singlet:
$\frac{1}{\sqrt{3}}\sum_{m}a_{im\uparrow}^{+}a_{im\downarrow}^{+}|0\rangle$,
where the spin-up and spin-down electrons are situated on a site
$i$ in the same quantum state $m$ (here $|0\rangle$ is the neutral
$\texttt{C}_{60}$ molecule, the quantum number $m$ labels the
three orthogonal states of $t_{1u}$ symmetry). The local singlet
state competes with the normal state (high spin state) of two
electrons $a_{im_{1}\uparrow}^{+}a_{im_{2}\uparrow}^{+}|0\rangle$
dictated by Hund's rule. Using this assumption some important
results have obtained. In a work \cite{suz} the density of states
in a band originated from $t_{1u}$ level has been calculated by
applying the unrestricted Hartree-Fock approximation and the many
body perturbation method. It has been found that
$\texttt{A}_{2}\texttt{C}_{60}$ and
$\texttt{A}_{4}\texttt{C}_{60}$ are nonmagnetic semiconductors and
the band gaps in these materials are cooperatively formed by the
electron-electron and electron-vibron interactions. On the other
hand, it is difficult to predict within the model whether the
following materials $\texttt{A}_{1}\texttt{C}_{60}$,
$\texttt{A}_{3}\texttt{C}_{60}$ and
$\texttt{A}_{5}\texttt{C}_{60}$ are metallic or not, it has been
concluded at least that the materials are on the border of the
metal-insulator transition. In the work \cite{gun1} it is
conjectured that the Mott-Hubbard transition takes place for
$U/W\sim\sqrt{N}$, where $N$ is an orbital degeneracy (for
alkali-doped fullerides $N=3$) due to the matrix elements for the
hoping of an electron or a hole from a site $i$ to a nearby site
$j$ are enhanced as $\langle
i|t_{ij}a^{+}_{im\sigma}a_{jm\sigma}|j\rangle=\sqrt{N}t$, thus the
degeneracy increases $U_{c}$ (critical value of the on-site
Coulomb repulsion such that if $U>U_{c}$ then a material is MI).
In the same time in \cite{koch} an analogous expression $\langle
i|\widehat{H}|j\rangle=\sqrt{k}t$ has been obtained, but the
factor $k$ depends on both the degeneracy and the filling $n$, so
for $n=3$ the enhancement is $k=1.73$, for $n=2,4$ - $k=1.57$, for
for $n=1,5$ - $k=1.21$. Thus the degeneracy contributes to the
metallization of the systems. However, as pointed above, the
conductivity depends on filling $n$ radically (for odd $n$ -
metals, for even $n$ - insulators), therefore the competition
between the Jahn-Teller effect and the Hund's rule coupling must
be accounted \cite{han1}.

In a work \cite{han2} it is shown that in
$\texttt{A}_{3}\texttt{C}_{60}$ the local pairing is crucial in
reducing the effects of the Coulomb repulsion. So, for the
Jahn-Teller $H_{g}$ phonons the attractive interaction is
overwhelmed by the Coulomb repulsion. Superconductivity remains,
however, even for $U_{vib}\ll U$, and $T_{c}$ drops surprisingly
slowly as $U$ is increased. The reason is as follows. For
noninteracting electrons the hopping tends to distribute the
electrons randomly over the molecular levels. This makes more
difficult to add or remove an electron pair with the same quantum
numbers $m$. However as $U$ is large $U>W$ the electron hopping is
suppressed and the local pair formation becomes more important.
Thus the Coulomb interaction actually helps the local pairing.
This leads to new physics in the strongly correlated low-bandwidth
solids, due to the interplay between the Coulomb and
electron-vibron interactions. In a such system the Eliashberg
theory breaks down because of the closeness to a metal-insulator
transition. Because of the local pairing, the Coulomb interaction
enters very differently for Jahn-Teller and non-Jahn-Teller
models, and it cannot be easily described by a Coulomb
pseudopotential $g-\mu^{\ast}$. Theoretical phase diagram for
$\texttt{A}_{3}\texttt{C}_{60}$ systems has been obtained with the
DMFT analysis in \cite{nom1,nom2}. There are three phase: the
superconducting (SC) phase at low temperature, the normal phase at
more high temperatures and the phase of paramagnetic MI at bigger
volume per $\texttt{C}_{60}^{3-}$. In the same time the dome shape
of the SC phase is absent in the theoretical diagram. In this
model the s-wave superconductivity is characterized by a order
parameter $\Delta=\sum_{m=1}^{3}\left\langle
a_{im\downarrow}a_{im\uparrow}\right\rangle$, which describes
intraorbital Cooper pairs for the $t_{1u}$ electrons, $m$ and $i$
are the orbital and site (=molecule) indices respectively (the
site index in $\Delta$ has been omitted, because $\Delta$ does not
depend on a site - the solution is homogenous in space).
Superconducting mechanism is that in such a system we have $U'>U$
($U'$ is interorbital repulsion and $U$ is intraorbital one) due
to the el.-vib. interaction. Interesting observation is that the
double occupancy $\left\langle
n_{m\uparrow}n_{m\downarrow}\right\rangle$ on each molecule
increases toward the Mott transition and it jumply increases in a
point of transition from the metal phase to the MI phase.
Conversely, the double occupancy $\left\langle
n_{m\uparrow}n_{m'\downarrow}\right\rangle$ and spin $S$ per
molecule decrease toward the Mott transition and they jumply
decrease (to $S=1/2$) in the point of transition from the metal
phase to the insulator phase. In the same time the local pairs are
not bipolarons because in metallic phase the polarons (polaron
band) are absent up to the Mott transition and stable electron
pairs are not observed above $T_{c}$.

The local pairing hypothesis is confirmed with quantum Monte Carlo
simulations of low temperature properties of the two-band Hubbard
model with degenerate orbitals \cite{koga}. It have been clarified
that the SC state can be realized in a repulsively interacting
two-orbital system due to the competition between the intra- and
interorbital Coulomb interactions: it must be $U<U'$ for this. The
s-wave SC state appears along the first-order phase boundary
between the metallic and paired Mott states in the paramagnetic
system. The exchange interaction $J$ destabilizes the SC state
additionally. In \cite{hosh} the phase portrait of
$\texttt{A}_{3}\texttt{C}_{60}$ has been obtained using the DMFT
in combination with the continuous-time quantum Monte Carlo
method. In the theoretical diagram there are the dome shaped SC
region and spontaneous orbital-selective Mott (SOSM) state, in
which itinerant and localized electrons coexist and it is
identified as JT metal by the authors. The SOSM state is
stabilized near the Mott phase, while SC appears in the lower-$U$
region. The transition into the SOSM and AFM phases is of the
first order. In the same time, as discussed above, the line of
phase transition from JT metal to conventional metal is absent in
the experimental phase diagram (the molecular distortions
gradually increase as lattice expands), hence the JT metal is not
a phase, unlike the MJT insulator which is separated from the
metal state by a line of the first kind phase transition. In the
theoretical phase diagram the characteristic vertical cutoff of
the SC and metallic phases (at low temperature) by MJT insulator
phase is absent or weakly expressed (a hysteresis behavior is
observed near the Mott transition point), the SC phase extends far
into the region of large $U/W$ where the SOSM phase occurs.

In the present work we are aimed to find conditions of formation
of the local pairs on fullerene molecules, then, based on the
local pairing hypothesis, considering the Coulomb correlations and
the JT-distortion of the molecules, to propose a general approach
to description of the properties of alkali-doped fullerides
$\texttt{A}_{n}\texttt{C}_{60}$
($\texttt{A}=\texttt{K},\texttt{Rb},\texttt{Cs}$, $n=1\ldots 5$),
namely to demonstrate mechanism of superconductivity of
$\texttt{A}_{3}\texttt{C}_{60}$, conductivity of
$\texttt{A}_{n}\texttt{C}_{60}$ with $n=1,3,5$ and insulation of
the materials with $n=2,4$ with the loss of antiferromagnetic
properties, and to obtain analytically the phase diagram of
$\texttt{A}_{3}\texttt{C}_{60}$ with $\texttt{fcc}$ structure
which should be close to the experimental phase diagram.

\section{Local pairing}\label{model}

\begin{figure}[h]
\includegraphics[width=8cm]{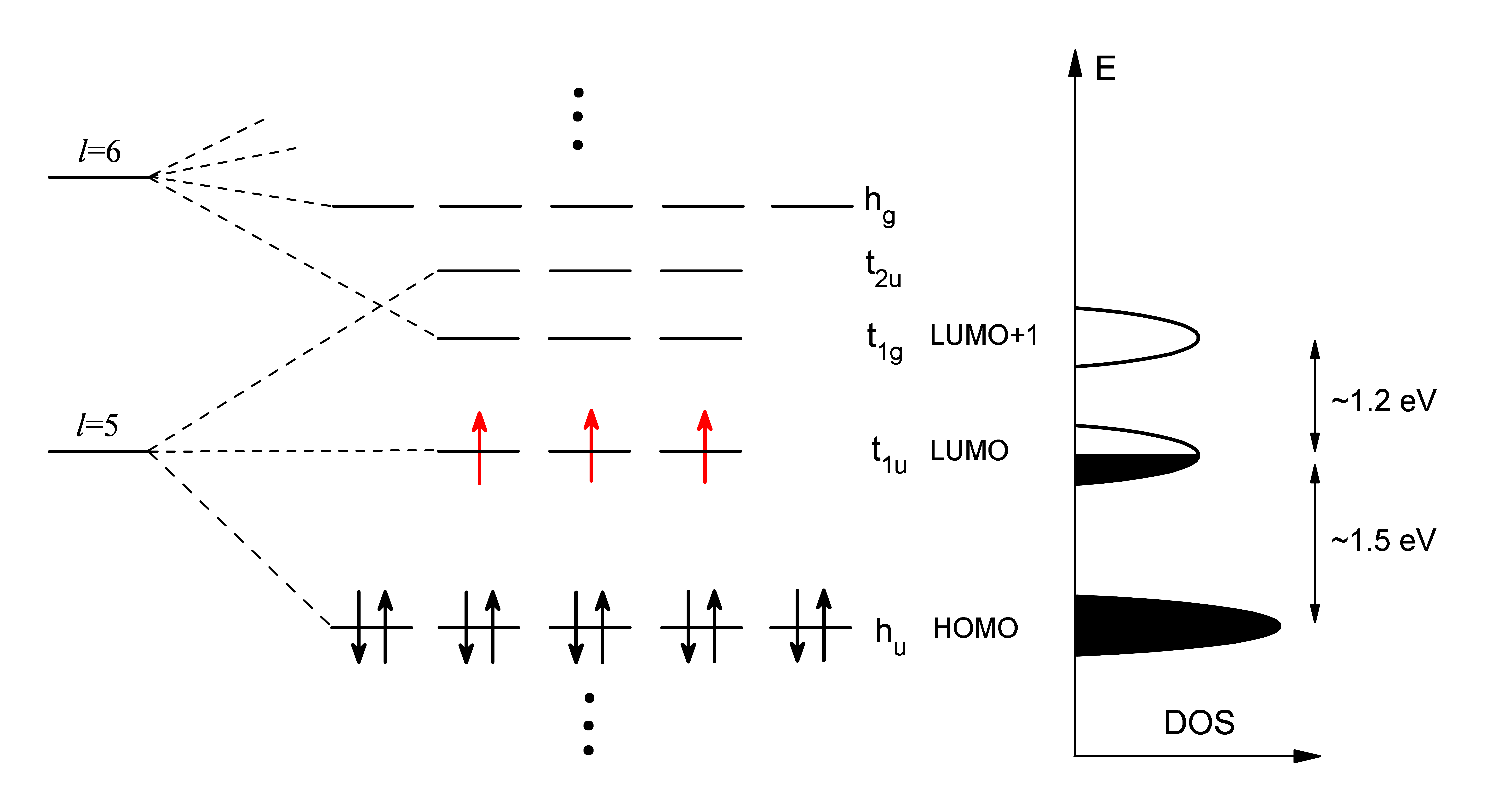}
\caption{The molecular levels of $\texttt{C}_{60}$ close to the
Fermi level. In a substance $\texttt{A}_{3}\texttt{C}_{60}$ the
atoms of alkali metal
$\texttt{A}=\texttt{K},\texttt{Rb},\texttt{Cs}$ give electrons to
the LUMO of the fullerene molecule (red color). Right panel:
corresponding band structure (density of state as a function of
energy $E$ schematically).} \label{Fig2}
\end{figure}

\begin{figure}[h]
\includegraphics[width=7cm]{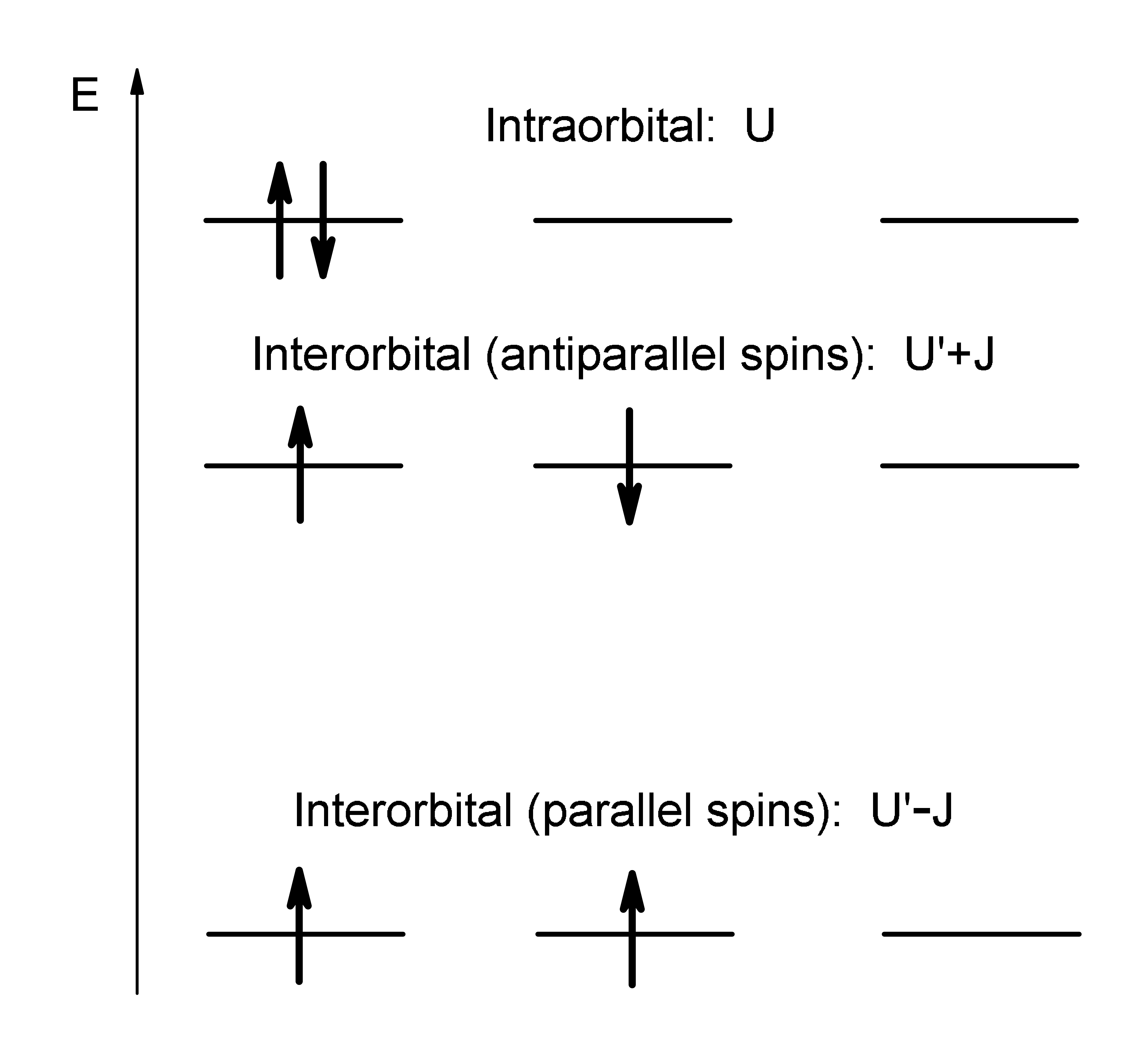}
\caption{The configurations of two electrons on the 3-fold
degenerate orbital. Ground state corresponds to configuration with
parallel spins. The intraorbital configuration has the largest
energy.} \label{Fig3}
\end{figure}

Due to the quasispherical structure of the molecule
$\texttt{C}_{60}$ the electron levels would be spherical harmonics
with the angular momentum $l$, however the icosahedral symmetry
generates the splits of the spherical states into icosahedral
representation \cite{sav,allon,hadd}. Fig.\ref{Fig2} shows
molecular levels close to the Fermi level. The LUMO is the 3-fold
degenerate $t_{1u}$ orbital (i.e. it can hold up to six
electrons). It is separated by about 1.5$\texttt{eV}$ from the
highest occupied molecular orbital (HOMO) and by about
1.2$\texttt{eV}$ from the next unoccupied level (LUMO+1). The
alkali-metal atoms give electrons to the empty $t_{1u}$ level so
that the level becomes partly occupied. Hamiltonian of the system
can be written in a form of three-orbital Hubbard Hamiltonian with
the Hund coupling \cite{koga,geor} and the electron-vibron
(Jahn-Teller phonons) interaction:
\begin{eqnarray}\label{1.1}
\widehat{H}&=&\sum_{ij}\sum_{m}\sum_{\sigma}\left(t_{ij}+(\varepsilon_{m}-\mu)\delta_{ij}\right)
a^{+}_{im\sigma}a_{jm\sigma}+U\sum_{i}\sum_{m}\sum_{\sigma}n_{im\uparrow}n_{im\downarrow}\nonumber\\
&+&\left(U'-J\right)\sum_{i}\sum_{m<m'}\sum_{\sigma}n_{im\sigma}n_{im'\sigma}
+\left(U'+J\right)\sum_{i}\sum_{m<m'}\sum_{\sigma}n_{im\sigma}n_{im'-\sigma}\nonumber\\
&+&V\sum_{\langle
ij\rangle}\sum_{m,m'}\sum_{\sigma\sigma'}n_{im\sigma}n_{jm'\sigma'}+\widehat{H}_{el-vib}+\widehat{H}_{vib},
\end{eqnarray}
where $a^{+}_{im\sigma}(a_{im\sigma})$ is the electron creation
(annihilation) operator in the orbital $m=1,2,3$ localized on the
site $i$; $n_{im\sigma}=a^{+}_{im\sigma}a_{im\sigma}$ is the
particle number operator; $\sigma(=\uparrow,\downarrow)$ is the
spin index; $t_{ij}$ is the hopping integral between neighboring
sites and the same orbitals (the bandwidth is $W=2z|t|$, where $z$
is the number of nearest neighbors), $t_{ii}=0$; $\varepsilon_{m}$
is the orbital energy; $\mu$ is the chemical potential,
$\delta_{ij}=1$ if $i=j$, $\delta_{ij}=0$ if $i\neq j$; the energy
$V$ is the Coulomb repulsion between neighboring sites ($i\neq
j$):
\begin{eqnarray}\label{1.1d}
  V = \int
  d\textbf{r}d\textbf{r}'|\phi_{m}^{i}(\textbf{r})|^{2}V_{c}^{ij}(\textbf{r},\textbf{r}')|\phi_{m'}^{j}(\textbf{r}')|^{2},
\end{eqnarray}
where $V_{c}^{ij}(\textbf{r},\textbf{r}')$ is a operator of
cross-site Coulomb interaction; $U$ is the intra-orbital on-site
Coulomb repulsion energy; $U'$ is the inter-orbital on-site
Coulomb repulsion energy; $J>0$ is the on-site exchange
interaction energy:
\begin{eqnarray}
  U &=& \int d\textbf{r}d\textbf{r}'|\phi_{m}(\textbf{r})|^{2}V_{c}(\textbf{r},\textbf{r}')|\phi_{m}(\textbf{r}')|^{2} \label{1.1a}\\
  U' &=& \int d\textbf{r}d\textbf{r}'|\phi_{m}(\textbf{r})|^{2}V_{c}(\textbf{r},\textbf{r}')|\phi_{m'}(\textbf{r}')|^{2} \label{1.1b}\\
  J &=& \int d\textbf{r}d\textbf{r}'\phi_{m}^{\ast}(\textbf{r})\phi_{m'}^{\ast}(\textbf{r}')
  V_{c}(\textbf{r},\textbf{r}')\phi_{m}(\textbf{r}')\phi_{m'}(\textbf{r})\label{1.1c},
\end{eqnarray}
where $V_{c}(\textbf{r},\textbf{r}')$ is a operator of on-site
Coulomb interaction. In a multi-orbital system the on-site Coulomb
interaction is presented with two terms: density-density
interaction energy $U$ ($U'$ for different orbitals) and the
energy of the double inter-orbital hoppings
$\sum_{ll'mm'}\sum_{\sigma
\sigma'}J_{l'm'lm}a^{+}_{il'\sigma}a^{+}_{im'\sigma'}a_{il\sigma}a_{im\sigma'}$
(where $l'\neq l,m'\neq m$) \cite{feng}. The hoppings' energy
cannot be reduced to the multiplication of operators of occupation
numbers like the density-density energy
$U'\sum_{lm}\sum_{\sigma\sigma'}n_{il\sigma}n_{im\sigma'}$.
However contribution of the exchange processes
$Ja^{+}_{im\sigma}a^{+}_{il\sigma'}a_{il\sigma}a_{im\sigma'}$ can
be reduced to the Hund coupling: variational calculations
demonstrate that Coulomb energy of two electrons in a singlet
state $|m_{1}\uparrow,m_{2}\downarrow\rangle$ is $U'+J$ and in a
triplet state $|m_{1}\uparrow,m_{2}\uparrow\rangle$ is $U'-J$. For
example, the first excited energy levels of a helium atom, which
are parallel and antiparallel spin configurations, are separated
from each other with energy gap $2J$. Physical reason of this
level splitting is that antisymmetric in spatial variables wave
function (parallel spins) of two electrons minimizes their Coulomb
repulsion due to maximal separation of the electrons in space in
comparison with state with the symmetric spatial part of wave
function (antiparallel spins) where the electrons are closer to
each other. This makes it possible to assume that the exchange
processes give main contribution in the interaction energy, and
contribution of other double inter-orbital hoppings can be
neglected. Thus in this approximation we can write Coulomb
interaction between electrons in a form of Hamiltonian (\ref{1.1})
where the effective density-density interaction is present only.
As a rule $U\gg J$, $U'\simeq U-2J$ (the RPA interaction
parameters for the family $\texttt{A}_{3}\texttt{C}_{60}$ taken
from \cite{nom3} are $U\sim
0.82\texttt{eV},U'\sim0.76\texttt{eV},J\sim31\texttt{meV}$). Then
$U'-J<U'+J<U$ that means Hund's rule: the electron configuration
with minimal energy has maximal spin. Corresponding electron
configurations for two electrons is shown in Fig.\ref{Fig3}.

In the $\texttt{C}_{60}^{n-}$ molecule the excess electrons are
coupled strongly to two $A_{g}$ and eight $H_{g}$ intramolecular
Jahn-Teller phonons (electron-vibron interaction). The operators
of the el.-vib. interaction and the vibrons' energy have forms
accordingly \cite{han1,han2,suz,gun,tak}:
\begin{eqnarray}
\widehat{H}_{el-vib}&=&\lambda_{A_{g}}\sum_{i}\sum_{M=1}^{2}\sum_{m=1}^{3}\sum_{m'=1}^{3}\sum_{\sigma}
V^{(0)}_{mm'}a^{+}_{im\sigma}a_{im'\sigma}\left(b^{+}_{i0M}+b_{i0M}\right)\nonumber\\
&+&\lambda_{H_{g}}\sum_{i}\sum_{\nu=1}^{5}\sum_{M=1}^{8}\sum_{m=1}^{3}\sum_{m'=1}^{3}\sum_{\sigma}
V^{(\nu)}_{mm'}a^{+}_{im\sigma}a_{im'\sigma}\left(b^{+}_{i\nu
M}+b_{i\nu M}\right),\label{1.2}\\
\widehat{H}_{vib}&=&\sum_{i}\sum_{M=1}^{2}\omega_{M}b^{+}_{i0M}b_{i0M}+\sum_{i}\sum_{\nu=1}^{5}\sum_{M=1}^{8}\omega_{M}b^{+}_{i\nu
M}b_{i\nu M},\label{1.2a1}
\end{eqnarray}
where $\lambda_{A_{g}}$ and $\lambda_{H_{g}}$ are coupling
constants to $A_{g}$ and $H_{g}$ intramolecular Jahn-Teller
phonons respectively, the elements $V^{(\nu)}_{mm'}$ of coupling
matrices $\widehat{V}^{\nu}$ are determined by icosahedral
symmetry. The matrix corresponding to $A_{g}$ phonons ($\nu=0$) is
\begin{equation}\label{1.2a}
    V^{(0)}=\left(%
\begin{array}{ccc}
  1 & 0 & 0 \\
  0 & 1 & 0 \\
  0 & 0 & 1 \\
\end{array}
\right).
\end{equation}
Coupling to $H_{g}$ phonons ($\nu=1\ldots 5$) is given by the
matrixes
\begin{eqnarray}\label{1.2b}
    &V^{(1)}=\frac{1}{2}\left(%
\begin{array}{ccc}
  -1 & 0 & 0 \\
  0 & -1 & 0 \\
  0 & 0 & 2 \\
\end{array}
\right)\quad V^{(2)}=\frac{\sqrt{3}}{2}\left(%
\begin{array}{ccc}
  1 & 0 & 0 \\
  0 & -1 & 0 \\
  0 & 0 & 0 \\
\end{array}
\right)\quad V^{(3)}=\frac{\sqrt{3}}{2}\left(%
\begin{array}{ccc}
  0 & 1 & 0 \\
  1 & 0 & 0 \\
  0 & 0 & 0 \\
\end{array}
\right)\nonumber\\
&V^{(4)}=\frac{\sqrt{3}}{2}\left(%
\begin{array}{ccc}
  0 & 0 & 1 \\
  0 & 0 & 0 \\
  1 & 0 & 0 \\
\end{array}
\right)\quad V^{(5)}=\frac{\sqrt{3}}{2}\left(%
\begin{array}{ccc}
  0 & 0 & 0 \\
  0 & 0 & 1 \\
  1 & 0 & 0 \\
\end{array}
\right).
\end{eqnarray}
Vibrational energies for the $A_{g}$ and $H_{g}$ modes are within
the limits $\omega=200\ldots 1800\texttt{cm}^{-1}$
\cite{gun0,koch}.

\begin{figure}[h]
\includegraphics[width=7cm]{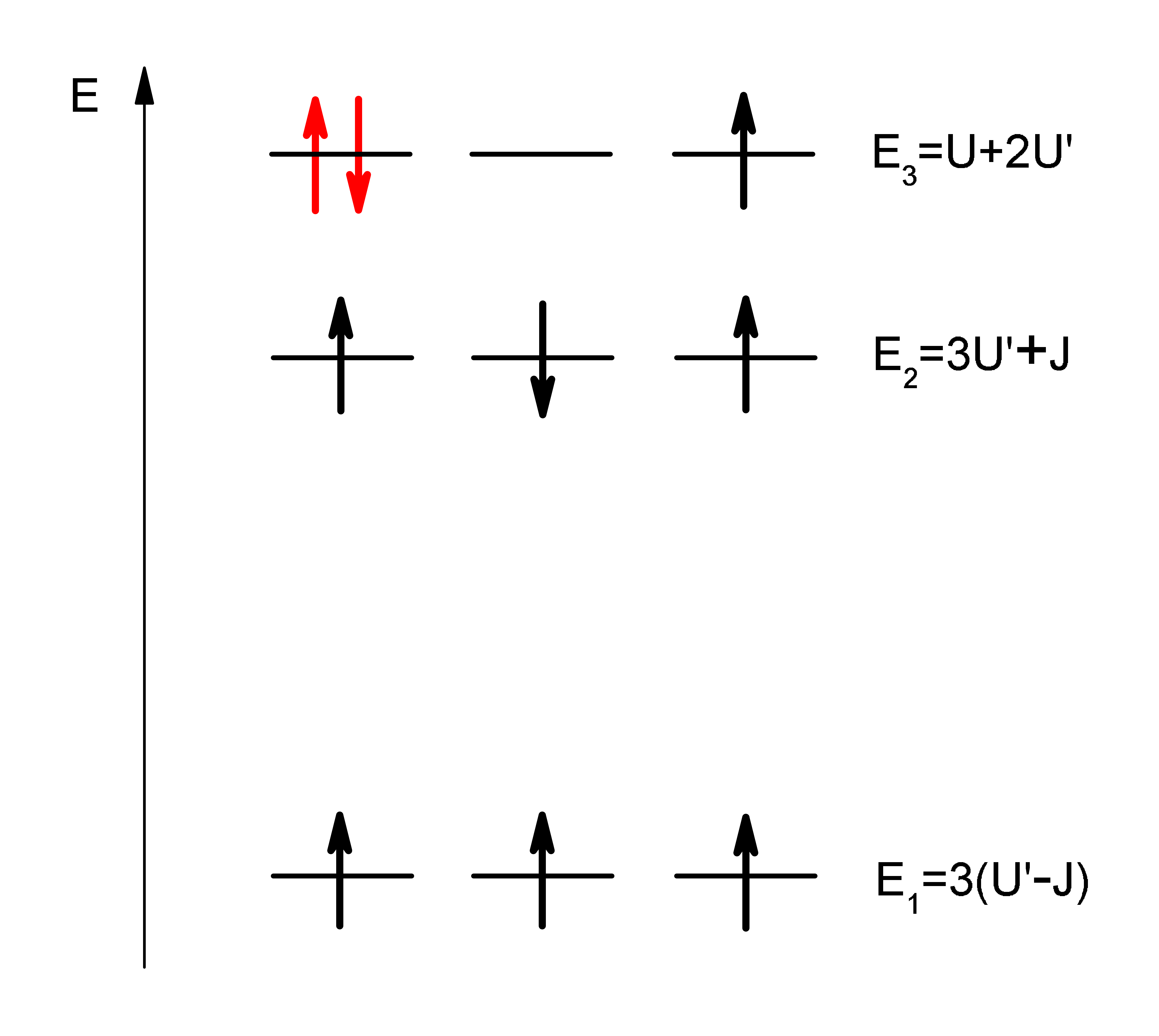}
\caption{The configurations of three electrons on LUMO of a
molecule $\texttt{C}_{60}$. Ground state corresponds to
configuration with parallel spins. The configuration with a local
pair (red color) has the largest energy.} \label{Fig4}
\end{figure}

According to the model of local pairing for superconductivity of
$\texttt{A}_{3}\texttt{C}_{60}$ the interaction of electrons with
$A_{g}$ and $H_{g}$ intramolecular oscillations favors the
formation of a local singlet \cite{han1,han2,lam1,lam2}:
\begin{equation}\label{1.4a}
\frac{1}{\sqrt{3}}\sum_{m}a_{m\uparrow}^{+}a_{m\downarrow}^{+}|0\rangle,
\end{equation}
where the spin-up and spin-down electrons have the same $m$
quantum number. Here $|0\rangle$ is the neutral $\texttt{C}_{60}$
molecule, the quantum number $m$ labels the three orthogonal
states of $t_{1u}$ symmetry. In contrast, the normal state (high
spin state) of two electrons is
\begin{equation}\label{1.4b}
\frac{1}{\sqrt{3}}\sum_{m_{1}<m_{2}}a_{m_{1}\uparrow}^{+}a_{m_{2}\uparrow}^{+}|0\rangle.
\end{equation}
As noted in \cite{han2}, for noninteracting electrons the hopping
tends to distribute the electrons randomly over the molecular
levels. However if the on-site Coulomb repulsion $U$ is large so
that $U>W$ the electron hopping is suppressed and the local pair
formation becomes more important. Thus Coulomb interaction
actually helps the local pairing and superconductivity is result
of interplay between the el.-vib. interaction, the Coulomb
blockade on a site and the hopping between neighboring sites.
Superconductivity is expected to exist in this material right up
to the Mott transition.

Based on the local pairing approach the Hamiltonian (\ref{1.1})
can be simplified in the following manner. Fig.\ref{Fig4} shows
energies for different electron configurations of
$\texttt{C}_{60}^{3-}$ molecule (that is for the molecule in
$\texttt{A}_{3}\texttt{C}_{60}$ solid). We can see Hund's rule:
the electron configuration with maximal spin $S=3/2$ has minimal
energy: $E_{3}-E_{1}=U-U'+3J\simeq 5J>E_{2}-E_{1}=4J>0$. Thus the
ground state of the system is a state $|m_{1}\sigma, m_{2}\sigma,
m_{3}\sigma\rangle$, in order to form a local pair (\ref{1.4a})
the energy $E_{3}-E_{1}=U-U'+3J$ must be expended. Therefore
\textit{to study only the local pairing} it is sufficient to
measure the Coulomb energy of the local pairing configuration from
the energy of the ground state (Hund's rule configuration
$|\uparrow\uparrow\uparrow\rangle$). Thus in this approach two
electrons in the state $|m\uparrow m\downarrow\rangle$ on a
fullerene molecule "interact" with energy $U-U'+3J$ which is the
Hund coupling, but Coulomb correlation effects between electrons
on neighboring molecules are neglected by the averaging. Then we
can reduce the Hamiltonian (\ref{1.1}) to a form:
\begin{eqnarray}\label{1.3}
\widehat{H}_{eff}&=&\frac{1}{2}\left(U'-J\right)\sum_{i}\langle
n_{i}\rangle\left(\langle n_{i}\rangle-1\right)+
V\sum_{\langle ij\rangle}\langle n_{i}\rangle\langle n_{j}\rangle\\
&+&\sum_{ij}\sum_{m}\sum_{\sigma}\left(t_{ij}+(\varepsilon_{m}-\mu)\delta_{ij}\right)
a^{+}_{im\sigma}a_{jm\sigma}
+\left(U-U'+3J\right)\sum_{i}\sum_{m}n_{im\uparrow}n_{im\downarrow}+\widehat{H}_{el-vib}+\widehat{H}_{vib},\nonumber
\end{eqnarray}
where $\langle n_{i}\rangle$ is an average occupation number of a
site $i$. Thus the effective Coulomb repulsion $U-U'+3J\sim
0.15\texttt{eV}$, which is much smaller than the on-site Coulomb
repulsion $U\sim 0.8\texttt{eV}$, resists to the local pairing,
unlike the usual Holstein-Hubbard model without degeneration.

\begin{figure}[h]
\includegraphics[width=12cm]{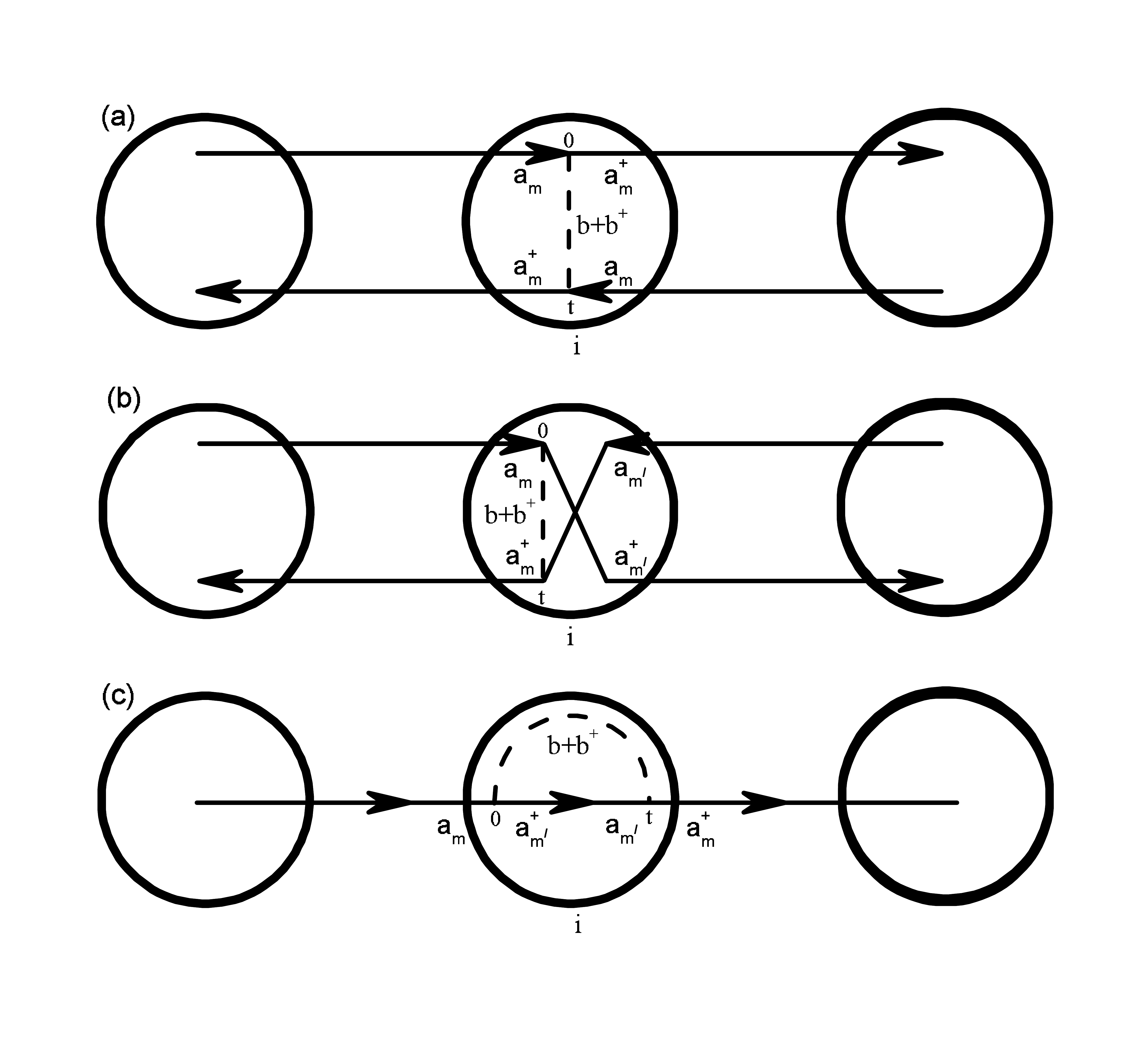}
\caption{interaction between two electrons on a site $i$ mediated
by the exchange of a vibron localized in this molecule: the case
(a) corresponds to a direct process
$\frac{2}{\omega_{0}}\sum_{\nu}\lambda^{2}_{\nu}V^{(\nu)}_{mm}V^{(\nu)}_{mm}
a^{+}_{im\uparrow}a_{im\uparrow}a^{+}_{im\downarrow}a_{im\downarrow}$,
the case (b) corresponds to an exchange process
$\frac{2}{\omega_{0}}\sum_{\nu}\lambda^{2}_{\nu}V^{(\nu)}_{mm'}V^{(\nu)}_{mm'}
a^{+}_{im\uparrow}a_{im'\uparrow}a^{+}_{im\downarrow}a_{im'\downarrow}$,
where $m\neq m'$. The picture (c) corresponds to the shift of
electron's energy due to the el.-vib. interaction on a site.}
\label{Fig5}
\end{figure}

We can eliminate the vibron variables using perturbation theory.
Let the molecular vibrations occur with a certain frequency
$\omega_{0}$. Since a vibron is localized on a molecule then
el.-el interaction mediated by the exchange of a vibron has a
form:
\begin{eqnarray}\label{1.3e1}
\widehat{H}_{el-el} &=&
\sum_{i}\sum_{m,m'}\sum_{\nu}\lambda^{2}_{\nu}V^{(\nu)}_{mm'}V^{(\nu)}_{mm'}
a^{+}_{im\uparrow}a_{im'\uparrow}a^{+}_{im\downarrow}a_{im'\downarrow}\nonumber\\
&&(-i)\int_{-\infty}^{+\infty} \left\langle
0\left|\widehat{T}\left\{\left(b_{\nu i}(t)+b_{\nu
i}^{+}(t)\right)\left(b_{\nu i}+b_{\nu
i}^{+}\right)\right\}\right|0\right\rangle dt,
\end{eqnarray}
where $\lambda_{0}=\lambda_{A_{g}}$,
$\lambda_{\nu}=\lambda_{H_{g}}$ for $\nu=1\ldots5$, $\widehat{T}$
is a Bose time-ordering operator,
$b(t)=e^{i\omega_{0}b^{+}bt}be^{-i\omega_{0}b^{+}bt}$ is an
annihilation (creation) phonon operator in Heisenberg
representation, $|0\rangle$ is a zero-phonon state. Expression
under integral is propagation of a vibron in a time $t$. The
corresponding processes are shown in Fig.(\ref{Fig5}). Then
\begin{eqnarray}\label{1.3e2}
\widehat{H}_{el-el} &=&
\sum_{i}\sum_{m,m'}\sum_{\nu}\lambda^{2}_{\nu}V^{(\nu)}_{mm'}V^{(\nu)}_{mm'}
a^{+}_{im\uparrow}a_{im'\uparrow}a^{+}_{im\downarrow}a_{im'\downarrow}\nonumber\\
&&(-i)\int_{-\infty}^{+\infty}
\left[\theta_{t}e^{-i\omega_{0}t}+\theta_{-t}e^{i\omega_{0}t}\right]dt=-\frac{2}{\omega_{0}}\sum_{i}\sum_{m,m'}\sum_{\nu}\lambda^{2}_{\nu}V^{(\nu)}_{mm'}V^{(\nu)}_{mm'}
a^{+}_{im\uparrow}a_{im'\uparrow}a^{+}_{im\downarrow}a_{im'\downarrow},
\end{eqnarray}
where $\theta_{t}=1$ if $t\geq 0$, $\theta_{t}=0$ if $t<0$. The
same result we have in frequency representation of the vibron's
propagator:
\begin{eqnarray}\label{1.3e3}
\widehat{H}_{el-el} &=&
\sum_{i}\sum_{m,m'}\sum_{\nu}\lambda^{2}_{\nu}V^{(\nu)}_{mm'}V^{(\nu)}_{mm'}
a^{+}_{im\uparrow}a_{im'\uparrow}a^{+}_{im\downarrow}a_{im'\downarrow}
\frac{-i}{2\pi\omega_{0}}\int_{-\infty}^{+\infty}\frac{2\omega_{0}}{\omega^{2}-\omega^{2}_{0}+2i\delta\omega_{0}}d\omega\nonumber\\
&=&-\frac{2}{\omega_{0}}\sum_{i}\sum_{m,m'}\sum_{\nu}\lambda^{2}_{\nu}V^{(\nu)}_{mm'}V^{(\nu)}_{mm'}
a^{+}_{im\uparrow}a_{im'\uparrow}a^{+}_{im\downarrow}a_{im'\downarrow}.
\end{eqnarray}
At nonzero temperature we should pass to complex time
$it=\rightarrow\tau$, where $\tau=-1/T\ldots 1/T$. This is the
same that to pass to complex frequency $\omega\rightarrow
i\omega_{n}$, where $\omega_{n}=\pi nT$. Then Eq.(\ref{1.3e3})
takes a form:
\begin{eqnarray}\label{1.3e4}
\widehat{H}_{el-el} &=&
-\sum_{i}\sum_{m,m'}\sum_{\nu}\lambda^{2}_{\nu}V^{(\nu)}_{mm'}V^{(\nu)}_{mm'}
a^{+}_{im\uparrow}a_{im'\uparrow}a^{+}_{im\downarrow}a_{im'\downarrow}
\frac{T}{\omega_{0}}\sum_{n=-\infty, n\neq 0}^{+\infty}\frac{2\omega_{0}}{(\pi nT)^{2}+\omega^{2}_{0}}\nonumber\\
&=&-\frac{2}{\omega_{0}}\left[\coth\left(\frac{\omega_{0}}{T}\right)-\frac{T}{\omega_{0}}\right]\sum_{i}\sum_{m,m'}
\sum_{\nu}\lambda^{2}_{\nu}V^{(\nu)}_{mm'}V^{(\nu)}_{mm'}
a^{+}_{im\uparrow}a_{im'\uparrow}a^{+}_{im\downarrow}a_{im'\downarrow}.
\end{eqnarray}
The addendum with $n=0$ is excluded from the sum, because it
corresponds to scattering of electron on the thermal phonons,
which do not influence on the pairing of electrons \cite{ginz}. We
can see that at $T\rightarrow 0$ Eq.(\ref{1.3e4}) goes into
Eq.(\ref{1.3e3}) as
$\frac{2\lambda^{2}}{\omega_{0}}\left(1-\frac{T}{\omega_{0}}\right)$,
that is at increasing of temperature the effectiveness of el.-vib.
interaction is decreasing. At $T\rightarrow\infty$
($T\gg\omega_{0}$) the el.-el. coupling via vibrons is
$\frac{2\lambda^{2}}{3T}\rightarrow 0$.

Thus we can see that the el.-el. interaction mediated by the
exchange of vibrons can be reduced to BCS-like interaction (point,
nonretarded) within single orbital $m$ that corresponds to
diagonal elements of the matrixes $\widehat{V}^{(\nu)}$:
\begin{equation}\label{1.3e5}
    U_{vib}^{mmmm}=\frac{2}{\omega_{0}}\left[\coth\left(\frac{\omega_{0}}{T}\right)-\frac{T}{\omega_{0}}\right]\sum_{\nu}\lambda^{2}_{\nu}V^{(\nu)}_{mm}V^{(\nu)}_{mm}
\end{equation}
and between different orbitals $m$ and $m'$ that corresponds to
the nondiagonal elements:
\begin{equation}\label{1.3e6}
    U_{vib}^{mm'mm'}=\frac{2}{\omega_{0}}\left[\coth\left(\frac{\omega_{0}}{T}\right)-\frac{T}{\omega_{0}}\right]\sum_{\nu}\lambda^{2}_{\nu}V^{(\nu)}_{mm'}V^{(\nu)}_{mm'}.
\end{equation}
For simplicity we suppose that the interaction energies are the
same for all orbitals $m$: $U_{vib}^{mmmm}=U_{vib}$ - direct
(intraorbital) interaction, $U_{vib}^{mm'mm'}=J_{vib}$ - exchange
(interorbital) interaction. It should be noticed that $J_{vib}>0$
as we can see using the matrixes (\ref{1.2a},\ref{1.2b}). In
addition we should consider the process
$\sum_{\nu}\lambda^{2}_{\nu}V^{(\nu)}_{mm}V^{(\nu)}_{m'm'}
a^{+}_{im\uparrow}a_{im\uparrow}a^{+}_{im'\downarrow}a_{im'\downarrow}$
which corresponds to the direct interorbital interaction
$U_{vib}'\propto
\sum_{\nu}\lambda^{2}_{\nu}V^{(\nu)}_{mm}V^{(\nu)}_{m'm'}$.
However, using the matrixes (\ref{1.2a},\ref{1.2b}) we can see
that $A_{g}$ and $H_{g}$ phonons give contribution to the
interaction $U_{vib}^{mmm'm'}$ with different signs:
$V^{0}_{mm}V^{0}_{m'm'}=1$,
$\sum_{\nu=1}^{5}V^{(\nu)}_{mm}V^{(\nu)}_{m'm'}=-1/2$, unlike the
intraorbital and the exchange interorbital interactions. Hence it
is possible to choose such values of the coupling constants
$\lambda_{A_{g}}$ and $\lambda_{H_{g}}$ so that $U_{vib}'=0$ (for
example, $\lambda_{H_{g}}=2\lambda_{A_{g}}$ for the model with the
same frequency $\omega_{0}$). Thus this interaction can be much
weaker than $U_{vib}$ and $J_{vib}$, that is confirmed by
numerical calculation for vibron-mediated interactions in
$\texttt{Cs}_{3}\texttt{C}_{60}$ \cite{nom4}. Then the Hamiltonian
(\ref{1.3}) takes a form:
\begin{eqnarray}\label{1.4}
\widehat{H}_{eff}&=&\sum_{ij}\sum_{m}\sum_{\sigma}\left(t_{ijmm}+(\varepsilon_{m}-\mu)\delta_{ij}\right)
a^{+}_{im\sigma}a_{jm\sigma}\nonumber\\
&+&\left(U-U'+3J-U_{vib}\right)\sum_{i}\sum_{m}n_{im\uparrow}n_{im\downarrow}
-J_{vib}\sum_{i}\sum_{m'\neq
m}a_{im\uparrow}^{+}a_{im'\uparrow}a_{im\downarrow}^{+}a_{im'\downarrow},
\end{eqnarray}
where we have omitted the constant contribution in energy (the
first and the second terms in Eq.(\ref{1.3})) and the vibrons'
energy $\widehat{H}_{vib}$. If each molecule is isolated, that is
$t_{ij}=0$, and the el.-vib. interaction is stronger than the Hund
coupling, that is $U-U'+3J-U_{vib}-2J_{vib}<0$, then the anti-Hund
distribution of electrons over orbitals occurs (the low-spin state
(\ref{1.4a})). Turning on the hopping $t_{ij}\neq 0$ between
molecules the electrons are collectivized and they aspire to
distribute randomly over the molecular levels. Then the local
pairing (\ref{1.4a}) is determined by presence of the anomalous
averages $\langle a_{im\downarrow}a_{im\uparrow}\rangle$ and
$\langle a_{im\uparrow}^{+}a_{im\downarrow}^{+}\rangle$, which are
determined self-consistently over the entire system. To find the
anomalous averages we should distinguish them in Eq.(\ref{1.4}),
then the Hamiltonian takes a following form:
\begin{eqnarray}\label{1.5}
\widehat{H}_{eff}&=&\sum_{ij}\sum_{mm'}\sum_{\sigma}\left(t_{ij}^{mm}+(\varepsilon_{m}-\mu)\delta_{ij}\right)
a^{+}_{im\sigma}a_{jm\sigma}\nonumber\\
&+&\sum_{i}\sum_{m}\left[\Delta_{m}^{+}a_{im\uparrow}a_{im\downarrow}
+\Delta_{m}a_{im\downarrow}^{+}a_{im\uparrow}^{+}\right]\nonumber\\
&+&(U_{vib}-U+U'-3J)\sum_{i}\sum_{m}\langle
a_{im\uparrow}^{+}a_{im\downarrow}^{+}\rangle\langle
a_{im\downarrow}a_{im\uparrow}\rangle+J_{vib}\sum_{i}\sum_{m'\neq
m}\langle a_{im'\uparrow}^{+}a_{im'\downarrow}^{+}\rangle\langle
a_{im'\downarrow}a_{im'\uparrow}\rangle
\end{eqnarray}
where
\begin{eqnarray}\label{1.6}
    \Delta_{m}&=&\frac{U_{vib}-U+U'-3J}{N}\sum_{i}\langle
a_{im\downarrow}a_{im\uparrow}\rangle+\frac{J_{vib}}{N}\sum_{i}\sum_{m'\neq
m}\langle a_{im'\downarrow}a_{im'\uparrow}\rangle\nonumber\\
\Delta_{m}^{+}&=&\frac{U_{vib}-U+U'-3J}{N}\sum_{i}\langle
a_{im\uparrow}^{+}a_{im\downarrow}^{+}\rangle+\frac{J_{vib}}{N}\sum_{i}\sum_{m'\neq
m}\langle a_{im'\uparrow}^{+}a_{im'\downarrow}^{+}\rangle
\end{eqnarray}
is the order parameter. $N$ is the number of lattice sites (number
of the molecules). Eq.(\ref{1.6}) means that the order parameter
in such system is determined by the local pairing. It should be
noticed that the condensates $\langle
a_{im\downarrow}a_{im\uparrow}\rangle$ for different orbitals $m$
have the same phase because $J_{vib}>0$. In this model the
exchange energy $3J$ and the difference $U-U'\sim 2J$ resists the
attraction energies $U_{vib},J_{vib}$. As indicated above the
exchange energy $J$ is much less than Coulomb repulsion $U,U'$.
Thus for the local pairing a weaker condition $U_{vib}\gtrsim 5J$
than $U_{vib}\gtrsim U$ must be satisfied. The average number of
electrons per site
\begin{equation}\label{1.7}
    \sum_{m}\langle n_{m}\rangle=\frac{1}{N}\sum_{i}\sum_{m}\sum_{\sigma}\langle
a_{im\sigma}^{+}a_{im\sigma}\rangle
\end{equation}
is determined by the position of the chemical potential $\mu$, and
$\langle\ldots\rangle=\Xi^{-1}Tr\ldots\exp(-\widehat{H}/T)$
denotes averaging procedure where $\Xi$ is a partition function.

\begin{figure}[h]
\includegraphics[width=16cm]{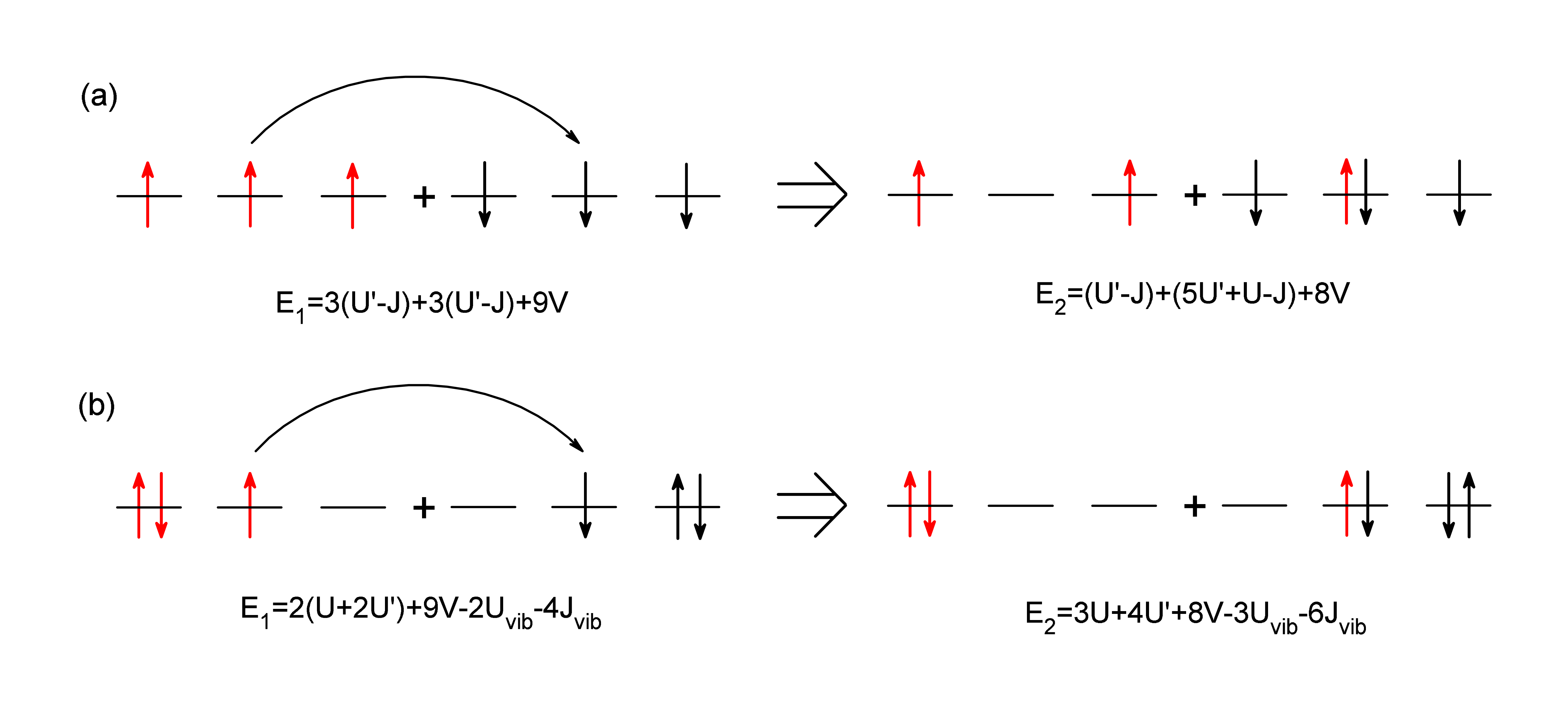}
\caption{The transfer of a charge from a site to a nearby site.
$E_{1}$ and $E_{2}$ are the energies of the electron
configurations before and after the transfer. (a) - the transfer
of an electron between sites with Hund's electron configuration
without el.-vib. interaction, (b) - the transfer of an electron
between sites with the local pairs configuration.} \label{Fig6}
\end{figure}

The effective Hamiltonian (\ref{1.3}) does not account Coulomb
correlations between electrons on neighboring sites, therefore to
study conduction or insulation of the material we should proceed
from the full Hamiltonian (\ref{1.1}). The on-site Coulomb
repulsions $U,U'$, the on-site exchange interaction energy $J$ and
the Coulomb repulsion between neighboring sites $V$ determine the
change of energy at transfer of an electron from a site to a
nearby site. This process is shown in Fig.\ref{Fig6}a. We can see
that the energy change in this process is
\begin{equation}\label{1.7a}
    \Delta E_{\texttt{Hund}}=E_{2}-E_{1}=U+4J-V,
\end{equation}
where $E_{1}$ is energy of an initial electron configuration (the
Hund's rule) of neighboring sites, $E_{2}$ is energy of the
configuration if to transfer an electron from the site to another.
A band conductor becomes insulator if an electron does not have
enough reserve of kinetic energy (which is a bandwidth $W$) to
overcome the Coulomb blockade on a site: $\Delta E> W$. In the
absence of long-range order all numerical and analytical
calculations indicate that the Mott-Hubbard transition should
occur in the region $0.5W<U_{c}<1.7W$ \cite{gebh}, so in
Hubbard-III-like analytical calculation of the superconducting
critical temperature in the presence of local Coulomb interactions
a critical value $U_{c}=W$ can be used \cite{rod}. Thus we can
assume that a band conductor becomes the Mott insulator if
\begin{equation}\label{1.7b}
    \frac{\Delta E}{W}\geqslant 1.
\end{equation}
For example, let us consider $\texttt{Rb}_{3}\texttt{C}_{60}$.
According to \cite{nom3}
$W=0.454\texttt{eV},U=0.92\texttt{eV},J=34\texttt{meV},V=0.27\texttt{eV}$,
then $\Delta E/W=1.73$. This means that
$\texttt{Rb}_{3}\texttt{C}_{60}$ would have to be the Mott
insulator. However $\texttt{Rb}_{3}\texttt{C}_{60}$ is a conductor
and, at low temperatures, is a superconductor even. As discussed
above, the results of \cite{gun1,koch} state that the degeneracy
and the filling contribute to the metallization of the systems due
to the matrix elements for the hopping of an electron or a hole
from a site $i$ to a nearby site $j$ are enhanced as $\langle
i|t_{ij}a^{+}_{im\sigma}a_{jm\sigma}|j\rangle=\sqrt{k}t$ where
$k>1$. However, it should be noted, that in a single-orbital
system the bandwidth $W$ is determined by the hopping as
$W=2z|t|$, where $z$ is number of the nearest neighbors. As stated
above in multi-orbital system the hopping is renormalized as
$t\rightarrow\sqrt{k}t$, hence the observed bandwidth $W$ has to
be determined with the renormalized hopping. Therefore the
degeneracy cannot change the criterium (\ref{1.7b}) which is
determined with the energetic balance. The degeneracy and the
filling essentially influence on the metal-insulator transition in
a multi-orbital system but due to configuration energy in $\Delta
E$ as will be demonstrated below.

Accounting of the el.-vib. interaction can change the situation.
As we can see from Eq.(\ref{1.4}) for 3-fold degenerate level
$t_{1u}$ each pair obtains energy $-U_{vib}-2J_{vib}$ due to
interaction within own orbital (for example $m=1$)
$1\leftrightarrow 1$ with energy $-U_{vib}$  and due to
interorbital coupling with other two orbitals
$1\leftrightarrow2,1\leftrightarrow 3$ with energy $-J_{vib}$ for
each. Formation of the local pair is possible if electron
configuration with the pair has energy which is less than energy
of electron configuration according to the Hund's rule (see
Fig.\ref{Fig4}): $U-U'+3J-U_{vib}-2J_{vib}<0$. Then the change of
energy at transfer of an electron from a site to the nearby site
is
\begin{equation}\label{1.7d}
    \Delta E_{\texttt{anti-Hund}}=E_{2}-E_{1}=U-V-U_{vib}-2J_{vib},
\end{equation}
as shown in Fig.\ref{Fig6}b. Here we can see more favorable
situation for conductivity because $\Delta E$ in this case is less
than in a case of the Hund's rule configuration (\ref{1.7a}) due
to configuration (exchange) energy $4J$ and the el.-vib.
interaction. However in the normal state (metallic) the anomalous
averages are absent: $\langle
a_{im\downarrow}a_{im\uparrow}\rangle=0$ and $\langle
a_{im\uparrow}^{+}a_{im\downarrow}^{+}\rangle=0$. This means the
local pair configuration is absent at the initial stage and we
have the transfer of an electron according to Hund's rule but with
el.-vib. interaction
\begin{equation}\label{1.7e}
    \Delta E_{\texttt{Hund}}=U+4J-V-U_{vib}-2J_{vib},
\end{equation}
where the local pair is formed on the second site -
Fig.\ref{Fig6}a. But $\Delta E_{\texttt{Hund}}>\Delta
E_{\texttt{anti-Hund}}$ due to configuration (exchange) energy
$4J$. Then it can be that $\Delta E_{\texttt{Hund}}/W>1$ and
$\Delta E_{\texttt{anti-Hund}}/W<1$. Hence transition from SC
state to the normal state would be transition to MI. However the
normal state is conducting due to following mechanism. Let $\Delta
E_{\texttt{Hund}}/W>1$ hence the hopping between molecules is
blocked and electrons are locked on a molecule. In the same time
$U-U'+3J-U_{vib}-2J_{vib}<0$, hence on the isolated molecule the
local pairing configuration is set. Then we obtain situation as in
Fig.\ref{Fig6}b where $\Delta E_{\texttt{anti-Hund}}/W<1$, that
allows an electron is transferred on the nearby molecule. This
makes the normal state to be conducting. Thus if the energy is
such that $\Delta E_{\texttt{anti-Hund}}/W<1$, then the material
becomes conductor and superconductor at low temperatures. In
Fig.\ref{Fig6}b we can see formation of configurations
$\texttt{C}_{60}^{2-}$ and $\texttt{C}_{60}^{4-}$ with zero spins
in the process of charge transferring. These configurations give a
gain in JT energy (all electrons are in the pairs with energy
$-U_{vib}-2J_{vib}$ each) but increase the Coulomb energy of a
crystal. Therefore these configurations would be formed within the
metal on very short time scales that do not imply static charge
segregation, that corresponds to the results of spin-lattice
relaxation measurements \cite{alloul1,alloul2}. As it will be
demonstrated in Section \ref{family} for the even systems, i.e.
$\texttt{A}_{n}\texttt{C}_{60}$ with $n=2,4$, the configurations
$\texttt{C}_{60}^{2-}$ and $\texttt{C}_{60}^{4-}$ are stabilized
that leads to insulation of these materials.

It should be noted that the local pair is not a local boson.
Following \cite{dzum1} if the size of a local pair $a_{p}$ is much
larger than the mean distance $R_{p}$ between the pairs then the
bosonization of such local pairs cannot be realized due to their
strong overlapping. Thus for fermionic nature of the pair it
should be $a_{p}/R_{p}>1$. Using the uncertainty principle, the
size of the local pair is defined as
$a_{p}(T)=\left(\frac{\hbar}{2|\Delta|}\right)\sqrt{\varepsilon_{F}/2m}$,
where the energy gap $|\Delta|$ plays role the binding energy in a
pair. The size is compared with $R_{p}=(3/4\pi n_{p})^{1/3}$,
where $n_{p}$ is the density of the pairs (half density of
particles which is determined by Fermi energy
$\varepsilon_{F}=\frac{\hbar^{2}}{2m}\left(\frac{3\pi^{2}N}{V}\right)^{2/3}\Rightarrow
R_{p}\sim\sqrt[3]{V/N}\sim \hbar/\sqrt{\varepsilon_{F}m}$). Then
we have
\begin{equation}\label{1.7f}
    \frac{a_{p}}{R_{p}}\sim\frac{\varepsilon_{F}}{|\Delta|}\sim\frac{W}{|\Delta|}\quad\begin{array}{cc}
      \gg 1 & \texttt{Cooper pairs}  \\
      \sim 1 & \texttt{crossover from BCS to BEC}  \\
      \ll 1 & \texttt{compact bosons}  \\
    \end{array}
\end{equation}
Maximal $T_{c}$ of alkali-doped fullerides is $35\texttt{K}$, then
$|\Delta|\approx 1.76T_{c}=62\texttt{K}\ll W\approx
0.5\texttt{eV}$. Thus the size of a pair is larger than average
distance electrons and the pair is \emph{smeared} over the
crystal. That is in the metallic phase the local pairs have
fermionic nature. At the border of transition to the MJT insulator
the bosonization of local pairs occurs due to Coulomb blockage of
hopping of electron between neighboring molecules. Thus electron
configuration with the local pairs is a \emph{dynamical}
configuration of all molecules unlike the statical one with a
compact local pair on each molecule for the MJT insulator state.

Since the local pairs smeared over the crystal the anomalous
averages $\langle a_{i\uparrow}a_{i\downarrow}\rangle$ exist in
momentum space too: $\langle
a_{-\textbf{k}\downarrow}a_{\textbf{k}\uparrow}\rangle$, and the
BCS-like theory can be applied for description of the
superconducting state of alkali-doped fullerides. Following a work
\cite{litak}, we can make transition from the site representation
(\ref{1.5}) into the reciprocal (momentum) space using relations:
\begin{equation}\label{1.8}
    a_{\textbf{k}m\sigma}=\frac{1}{\sqrt{N}}\sum_{j}e^{i\textbf{kr}_{j}}a_{jm\sigma},
    \quad
    a_{jm\sigma}=\frac{1}{\sqrt{N}}\sum_{j}e^{-i\textbf{kr}_{j}}a_{\textbf{k}m\sigma},
\end{equation}
then we obtain the effective Hamiltonian like Hamiltonian of a
multi-band superconductor:
\begin{eqnarray}\label{1.9}
\widehat{H}_{eff}&=&\sum_{m}\sum_{\textbf{k}}\sum_{\sigma}\xi_{m}(k)
a^{+}_{\textbf{k}m\sigma}a_{\textbf{k}m\sigma}\nonumber\\
&+&\sum_{\textbf{k}}\sum_{m}\left[\Delta_{m}^{+}a_{\textbf{k}m\uparrow}a_{-\textbf{k}m\downarrow}
+\Delta_{m}a_{-\textbf{k}m\downarrow}^{+}a_{\textbf{k}m\uparrow}^{+}\right],
\end{eqnarray}
with $\xi_{m}(k)=\varepsilon_{m}(k)-\mu$, and two last terms in
Eq.(\ref{1.5}) have been omitted as a constant. The homogeneous
equilibrium gaps are defined as
\begin{eqnarray}\label{1.10}
    \Delta_{m}&=&\frac{U_{vib}-U+U'-3J}{N}\sum_{\textbf{k}}\langle
a_{-\textbf{k}m\downarrow}a_{\textbf{k}m\uparrow}\rangle+\frac{J_{vib}}{N}\sum_{\textbf{k}}\sum_{m'\neq
m}\langle a_{-\textbf{k}m'\downarrow}a_{\textbf{k}m'\uparrow}\rangle\nonumber\\
\Delta_{m}^{+}&=&\frac{U_{vib}-U+U'-3J}{N}\sum_{\textbf{k}}\langle
a_{\textbf{k}m\uparrow}^{+}a_{-\textbf{k}m\downarrow}^{+}\rangle+\frac{J_{vib}}{N}\sum_{\textbf{k}}\sum_{m'\neq
m}\langle
a_{\textbf{k}m'\uparrow}^{+}a_{-\textbf{k}m'\downarrow}^{+}\rangle,
\end{eqnarray}
and the average number of electrons per site is
\begin{equation}\label{1.11}
    \sum_{m}\langle n_{m}\rangle=\frac{1}{N}\sum_{\textbf{k}}\sum_{m}\sum_{\sigma}\langle
a_{\textbf{k}m\sigma}^{+}a_{\textbf{k}m\sigma}\rangle
\end{equation}
Equations (\ref{1.10}) and (\ref{1.11}) should be solved
self-consistently. It is easy to find that
\begin{equation}\label{1.12}
   \langle
a_{-\textbf{k}m\downarrow}a_{\textbf{k}m\uparrow}\rangle=\frac{\Delta_{m}}{2E_{m}}\tanh\frac{E_{m}}{2T}
\end{equation}
and
\begin{equation}\label{1.13}
    \langle
a_{\textbf{k}m\sigma}^{+}a_{\textbf{k}m\sigma}\rangle=\frac{1}{2}\left(1-\frac{\xi_{m}(k)}{E_{m}}\tanh\frac{E_{m}}{2T}\right),
\end{equation}
where $E_{m}=\sqrt{\xi_{m}(k)+\Delta_{m}^{2}}$. Thus a system with
the local pairing $\langle a_{im\downarrow}a_{im\uparrow}\rangle$
is equivalent to a multi-band superconductor with a continual
pairing $\langle
a_{-\textbf{k}m\downarrow}a_{\textbf{k}m\uparrow}\rangle$ in each
band. The multi-band theory \cite{asker1,asker2,asker3} can be
mapped onto an effective single-band problem \cite{grig2}. Thus
the three-orbitals $t_{1u}$ system can be reduced to an effective
single-band superconductor, the more so that the condensates
$\langle a_{-\textbf{k}m\downarrow}a_{\textbf{k}m\uparrow}\rangle$
for different orbitals $m$ have the same phase because
$J_{vib}>0$. If to suppose the dispersion law of electrons in
$t_{1u}$ conduction band $\xi_{m}(k)$ is the same for all orbitals
$m$: $\xi_{1}=\xi_{2}=\xi_{3}=\xi$, then from Eq.(\ref{1.10}) we
obtain a simple equation for critical temperature:
\begin{eqnarray}\label{1.13a}
&&\left|\begin{array}{ccc}
  \left(U_{vib}-U+U'-3J\right)\zeta-1 & J_{vib}\zeta & J_{vib}\zeta \\
  J_{vib}\zeta & \left(U_{vib}-U+U'-3J\right)\zeta-1 & J_{vib}\zeta \\
  J_{vib}\zeta & J_{vib}\zeta & \left(U_{vib}-U+U'-3J\right)\zeta-1 \\
\end{array}\right|=0\nonumber\\
&&\nonumber\\
&&\Longrightarrow\left(U_{vib}+2J_{vib}-U+U'-3J\right)\zeta=1,
\end{eqnarray}
where
\begin{equation}\label{1.13b}
    \zeta=\frac{1}{N}\sum_{\textbf{k}}\frac{1}{2\xi(k)}\tanh\frac{\xi(k)}{2T_{c}}
\end{equation}
In the alkali doped fulerides the conduction band is narrow
$W=0.3\ldots 0.5\texttt{eV}$ and energies of vibrons are large
$\omega\sim 0.2eV$, thus  almost all electrons take part in
el.-vib. interaction ($2\omega\sim W$), unlike electrons in
conventional metals where only electrons near Fermi surface
interact via phonons because $\omega\ll W$ there. Hence the final
results are weakly sensitive to distribution of the density of
states in conduction band, that is some averaged density can be
used in this case, unlike conventional conductors where the
results strongly depend on the density on Fermi level. Thus we can
suppose the density of states in the conduction band is a constant
$\nu=\nu_{0}$ if $-W/2<\xi<W/2$, otherwise $\nu=0$. Since
$\sum_{\textbf{k}}\langle n(\textbf{k})\rangle=V\int\nu(\xi)d\xi$
it can be seen from Eq.(\ref{1.11}) that
\begin{equation}\label{1.14}
    3=\sum_{m}\langle
    n_{m}\rangle=2\frac{V}{N}\sum_{m}\int_{-W/2}^{0}\nu_{0}d\xi\Longrightarrow\nu_{0}=\frac{N}{V}\frac{1}{W}.
\end{equation}
Then Eq.(\ref{1.13a}) is reduced to a form
\begin{equation}\label{1.15}
    1=(g-\mu_{c})\int_{-\Omega}^{\Omega}\frac{1}{2\xi}\tanh\frac{\xi}{2T_{c}}d\xi,
\end{equation}
where the integration is restricted by energy $\Omega=\min
(\omega_{0},W/2)$ proceeding from the rectangular approximation
for the density of states in conduction band, narrowness of
conduction band and large vibron's energy, for example for
$\texttt{Cs}_{3}\texttt{C}_{60}$ at normal pressure the bandwidth
is $W=0.341\texttt{eV}<2\omega_{0}\approx 0.4\texttt{eV}$. The
coupling constant $g$ is determined with el.-vib. interaction
$U_{vib}+2J_{vib}$, and the Coulomb pseudopotential $\mu_{c}$ is
determined with the Hund coupling $U-U'+3J$:
\begin{equation}\label{1.16}
    g=\frac{U_{vib}+2J_{vib}}{W}=\frac{\widetilde{U}_{vib}}{W}, \quad \mu_{c}=\frac{U-U'+3J}{W}.
\end{equation}
The therm $\widetilde{U}_{vib}\equiv U_{vib}+2J_{vib}$ corresponds
to the energy of attraction in a pair as discussed above. The
bandwidth of alkali-doped fullerides is $W\sim 0.5\texttt{eV}$,
the energy gap is $\Delta\approx 2T_{c}\sim30\ldots
60\texttt{K}\ll \varepsilon_{F}=W/2$ that is a change of the
chemical potential at transition to SC state can be neglected
unlike the systems with the strong attraction and low particle
density, where it can be $\Delta>\varepsilon_{F}$ and the change
of the chemical potential plays important role in formation of SC
state \cite{lok}. On the other hand the vibrational energies for
the $A_{g}$ and $H_{g}$ modes are $\omega_{0}\sim
0.2\texttt{eV}\sim \varepsilon_{F}$ that means Tolmachev's
weakening of the Coulomb pseudopotential by a factor
$\ln\frac{\varepsilon_{F}}{\omega}$ does not take place.

It should be noticed that the effect of weak renormalization of
electron's mass due to el.-vib. interaction, which is similar to
the effect in ordinary metals due to el.-phon. interaction, should
be absent in alkali-doped fullerides. As shown in \cite{jones} the
renormalization is consequence of el.-phon. interaction and of the
fact that the total momentum of the system
\begin{equation}\label{1.3c}
    \textbf{P}^{\texttt{tot}}=\textbf{p}+\sum_{\textbf{q}}b_{\textbf{q}}^{+}b_{\textbf{q}}\hbar\textbf{q}
\end{equation}
is a constant of the motion. Here $\textbf{p}$ is the momentum of
an electron, $\textbf{q}$ is the wave vector of a phonon. The
electron energy is renormalized as
\begin{equation}\label{1.3d}
    E=E_{0}-\alpha\omega_{0}+\frac{p^{2}}{2m}\frac{1}{1+\alpha/6},
\end{equation}
where $\alpha\sim\frac{\lambda^{2}}{\omega_{0}^{2}}$. However in
alkali-doped fullerides electrons interact with internal
oscillations (vibrons) of the fullerene molecules. Each vibron is
localized on own molecule and cannot propagate to nearby
molecules, unlike phonons in metals which propagate throw the
system as waves. This means that the wave vector $\textbf{q}$ is
not quantum number for vibrons and Eq.(\ref{1.3c}) does not have
physical sense. Then the electron's momentum $\textbf{p}$ should
conserves separately. Moreover unlike conventional metals, where
$U/W\ll 1$, the alkali-doped fullerides are strongly correlated
systems, where $U/W\sim 1$, therefore the renormalization of the
density of state is caused by the Coulomb correlations at presence
of the el.-vib. interaction, that will be discussed in the next
section.

In the same time in our case the shift of the electron's energy
$-\alpha\omega_{0}$ is a consequence of a process of el.-vib.
interaction on a cite like the el.-el. interaction via vibrons
considered above. This process is shown in Fig.(\ref{Fig5}c).
Since a vibron is localized on a molecule we can write the energy
shift using Eq.(\ref{1.2}) and the averaging analogously to
Eqs.(\ref{1.3e1},\ref{1.3e3}):
\begin{eqnarray}\label{1.16a}
\widehat{H}_{el.-vib.} &=&
\sum_{i}\sum_{m,m'}\sum_{\nu}\lambda^{2}_{\nu}V^{(\nu)}_{mm'}V^{(\nu)}_{mm'}
a^{+}_{im\uparrow}a_{im\uparrow}
\frac{i}{2\pi}\int_{-\infty}^{+\infty}\frac{2\omega_{0}}{\omega^{2}-\omega^{2}_{0}+2i\delta\omega_{0}}\frac{1}{\varepsilon-\omega-\varepsilon_{m'}+i\delta}
d\omega\nonumber\\
&=_{\varepsilon=\varepsilon_{m'}}&-\sum_{i}\sum_{m,m'}\sum_{\nu}\frac{\lambda^{2}_{\nu}}{\omega_{0}}V^{(\nu)}_{mm'}V^{(\nu)}_{mm'}
a^{+}_{im\uparrow}a_{im\uparrow}.
\end{eqnarray}
Expressions under the integral are propagator of a vibron and
propagator of an electron located on a molecular level
$\varepsilon_{m}$ (if we suppose $\varepsilon_{m}>0$ then
$\delta\rightarrow +0$, if $\varepsilon_{m}<0$ then
$\delta\rightarrow -0$). After calculation of the integral we have
switched to the mass surface $\varepsilon=\varepsilon_{m}$. Thus
interaction of an electron with the vibron field on a molecule
shifts the electron's energy as
\begin{eqnarray}\label{1.16b}
-\sum_{m'}\sum_{\nu}\frac{\lambda^{2}_{\nu}}{\omega_{0}}V^{(\nu)}_{mm'}V^{(\nu)}_{mm'},
\end{eqnarray}
whose modulus is the binding energy of an electron with
deformation of a molecule if the electron is localized on it.

\section{Phase diagram}\label{phase}

\subsection{Influence of Coulomb correlations}

The BCS formula (\ref{1.15}) do not account Coulomb correlations
between electrons on neighboring sites because it has been
obtained using the effective Hamiltonian (\ref{1.4}). For studying
of these phenomena we should proceed from the full Hamiltonian
(\ref{1.1}). The correlations are determined by the fact that in
order to transfer an electron from a site to a nearby site we have
to change the energy of the configuration as in Fig.\ref{Fig6}b.
On the other hand the possibility of this transferring determines
metallic properties of the material -
Eqs.(\ref{1.7a},\ref{1.7b},\ref{1.7d}). The Coulomb correlations
can be accounted with Gutzwiller-Brinkman-Rice approach
\cite{brink} by the following manner. Quasiparticle states are
determined with the pole expression of an electron propagator:
\begin{equation}\label{1.17}
    G(\textbf{k},\varepsilon)=\frac{Z}{\varepsilon-\xi(k)+i\gamma_{k}},
\end{equation}
where $0<Z\leq 1$ is the discontinuity in the single-particle
occupation number $n_{\textbf{k}}$ at the Fermi surface, in other
hand the function $Z$ determines intensity of the quasiparticle
peak at $\omega=0$ of a spectral function
$A(\omega)=\sum_{\textbf{k}}A(\textbf{k},\omega)$, where
$A(\textbf{k},\omega)=-\frac{1}{\pi}\texttt{Im}G(\textbf{k},\omega)\approx\frac{Z}{\pi}\frac{\gamma_{k}}{(\omega-\xi(k))^{2}+\gamma_{k}^{2}}$.
For noninteracting electrons $Z=1$. At transition from metal state
to MI state the peak disappears and in its place an energetic gap
appears separating two Hubbard subbands. The vanishing of $Z$
therefore marks the metal-insulator transition \cite{bulla}. Such
Coulomb correlations redistribute the density of state in a
conduction band. However in alkali-doped fullerides the el.-vib.
interaction takes the whole bandwidth $2\omega_{0}\sim W$ hence
the effective density of state is determined by the bandwidth
$\nu_{F}\sim 1/W$ up to the appearance of the Hubbard subbands. In
the Brinkman-Rice approach the renormalization parameter $Z$ is
obtained as
\begin{equation}\label{1.20}
    Z=1-\left(\frac{\Delta E}{\Delta E_{c}}\right)^{2}=1-\left(\frac{U-V-\widetilde{U}_{vib}}{W}\right)^{2},
\end{equation}
where $\Delta E_{c}$ is a critical energy change at transfer of an
electron from a site to a nearby site. The last part of this
formula is written for our model -
Eqs.(\ref{1.7a},\ref{1.7b},\ref{1.7d}). Then the condition
(\ref{1.7b}) is $Z\leq 0$. Using the propagator (\ref{1.17}) we
can obtain normal and anomalous Green functions from Gorkov's
equation:
\begin{eqnarray}
    G(\varepsilon_{n},\xi)=i\frac{Z\cdot\left(i\varepsilon_{n}+\xi\right)}
    {(i\varepsilon_{n})^{2}-\xi^{2}-Z^{2}|\Delta|^{2}}\label{1.21a}\\
    F(\varepsilon_{n},\xi)=i\frac{Z^{2}\Delta}
    {(i\varepsilon_{n})^{2}-\xi^{2}-Z^{2}|\Delta|^{2}},\label{1.21b}
\end{eqnarray}
where $\varepsilon_{n}=\pi T(2n+1)$. Then the self-consistency
condition for the order parameter is
\begin{equation}\label{1.22}
    \Delta=(g-\mu)
    T\sum_{n=-\infty}^{\infty}\int_{-\Omega}^{\Omega}d\xi iF(\varepsilon_{n},\xi)
    \Longrightarrow
    1=(g-\mu)Z^{2}\int_{-\Omega}^{\Omega}\frac{1}{2\xi}\tanh\frac{\xi}{2T_{c}}d\xi,
\end{equation}
where $\Omega=\min(\omega_{0},W/2)$ as discussed above. This
formula for the critical temperature $T_{c}$, unlike the formula
(\ref{1.15}), accounts the Coulomb correlations through the
renormalization parameter $Z$. For uncorrelated metal $Z=1$, at
the transition to MI $Z=0$. Thus Coulomb correlations suppress the
critical temperature.

\begin{table}[h]
\caption{Calculated in \cite{nom3} bandwidth $W$, cRPA screened
Coulomb parameters $U$, $U'$, $J$, $V$ and dielectric constant
$\varepsilon$ for the compounds with $\texttt{fcc}$-lattice:
$\texttt{K}$, $\texttt{Rb}$ and $\texttt{Cs}$ in superconducting
phases with maximum $T_{c}$ (at pressure $7\texttt{kbar}$), in the
vicinity of the metal-insulator transition ($2\texttt{kbar}$) and
in the antiferromagnetic insulating phase (normal pressure),
respectively. At the bottom of the table the Coulomb barrier
$U-U'+3J$ for the local pairing is calculated.}\label{tab1}
\begin{center}
\begin{tabular}{|c|ccccc|}
  \hline\rule{0cm}{0.5cm}
    & $K_{3}\texttt{C}_{60}$ & $\texttt{Rb}_{3}\texttt{C}_{60}$ & $\texttt{Cs}_{3}\texttt{C}_{60}$ (7\texttt{kbar}) & $\texttt{Cs}_{3}\texttt{C}_{60}$ (2\texttt{kbar}) & $\texttt{Cs}_{3}\texttt{C}_{60}$  \\[0.1cm]
  \hline\rule{0cm}{0.45cm}
  volume per $\texttt{C}_{60}$ $ (\texttt{A}^{3})$ & 722 & 750 & 762 & 784 & 804 \\
  \rule{0cm}{0.45cm}
  $W(\texttt{eV})$ & 0.502 & 0.454 & 0.427 & 0.379 & 0.341 \\
  \rule{0cm}{0.45cm}
  $U(\texttt{eV})$ & 0.82 & 0.92 & 0.94 & 1.02 & 1.08 \\
  \rule{0cm}{0.45cm}
  $U'(\texttt{eV})$ & 0.76 & 0.85 & 0.87 & 0.94 & 1.00 \\
  \rule{0cm}{0.45cm}
$J(\texttt{meV})$ & 31 & 34 & 35 & 35 & 36 \\
  \rule{0cm}{0.45cm}
  $V(\texttt{eV})$ & 0.24 $-$ 0.25 & 0.26 $-$ 0.27 & 0.27 $-$ 0.28 & 0.28 $-$ 0.29 & 0.30 \\
  \rule{0cm}{0.45cm}
  $\varepsilon_{\texttt{cRPA}}$ & 5.6 & 5.1 & 4.9 & 4.6 & 4.4 \\
\rule{0cm}{0.45cm}
  $U-U'+3J (\texttt{eV})$ & 0.153 & 0.172 & 0.175 & 0.185 &  0.188\\[0.1cm]
\hline
\end{tabular}
\end{center}
\end{table}

The RPA screened parameters of Coulomb interactions $U,U',J,V$ for
the family $\texttt{A}_{3}\texttt{C}_{60}$ have been calculated in
\cite{nom3} by the constrained RPA method (cRPA). The results are
presented in Tab.\ref{tab1}. The unscreened (bare) parameters are
$U=3.27\texttt{eV},U'=3.08\texttt{eV},J=96\texttt{meV},V=1.31\texttt{eV}$.
From the table we can see that the Coulomb parameters are
functions of volume per $\texttt{C}_{60}$ molecule: the screening
becomes less effective when the lattice is expanded and the
parameters aspire to the bare values. The screening can be
effectively described with the dielectric constant defined as
\cite{nom3}:
\begin{equation}\label{1.23a}
    \varepsilon_{\texttt{cRPA}}=\lim_{\textbf{Q}\rightarrow 0}\lim_{\omega\rightarrow 0}
    \frac{1}{\varepsilon_{\texttt{cRPA}}^{-1}\left(\textbf{q},\omega\right)},
\end{equation}
with $\omega$ being the frequency and
$\textbf{Q}=\textbf{q}+\textbf{G}$, where $\textbf{q}$ is a wave
vector in the first Brillouin zone and $\textbf{G}$ is a
reciprocal lattice vector. Thus the screened Coulomb interaction
and the bare one are connected as
\begin{equation}\label{1.23}
    U_{\texttt{cRPA}}\simeq\frac{U_{\texttt{bare}}}{\varepsilon_{\texttt{cRPA}}}.
\end{equation}
Here the static dielectric constant (\ref{1.23a}) can be used
since the plasmon energy in $\texttt{A}_{3}\texttt{C}_{60}$ is
$\sim 0.5\texttt{eV}$ \cite{gold} and the vibron energies are
$0.034\texttt{eV}\ldots 0.196\texttt{eV}$. Indeed, calculation in
\cite{nom1} shows the results of $U(\omega)$, $U’(\omega)$,
$J(\omega)$ are almost flat in the frequency region where the
vibron-mediated interactions $U_{vib}(\omega)$ and
$J_{vib}(\omega)$ are active ($\omega\lesssim 0.2 eV$).
Furthermore, the values of the cRPA interactions differ by less
than $15\%$ from the $\omega=0$ value up to, at least, $\omega=3
\texttt{eV}$ which is much larger than the bandwidth $\sim
0.5\texttt{eV}$.

\begin{figure}[h]
\includegraphics[width=10cm]{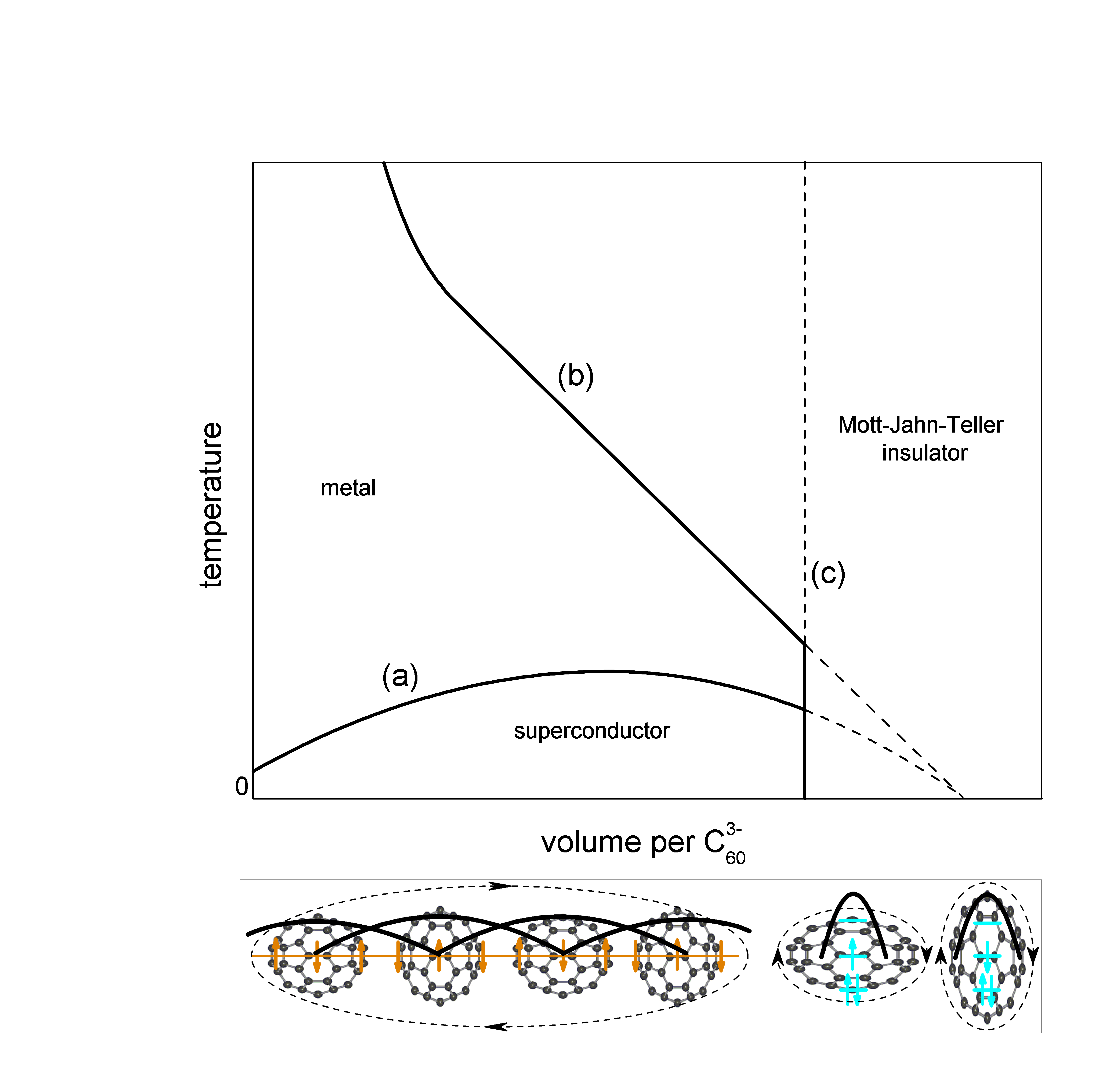}
\caption{Theoretical phase diagram of alkali-doped fullerides as a
function of volume per $\texttt{C}_{60}$. A line (a) is the
critical temperatures $T_{c}$ described with Eq.(\ref{1.22}). The
line separates the SC phase and the normal metallic phase. A line
(b) is the critical temperatures $T_{\texttt{MI}}$ described with
Eq.(\ref{1.25}). The line separates the normal metallic phase and
the MI phase. A line (c), described with Eq.(\ref{1.31}),
separates the metallic phase (normal and superconducting) and the
MJT insulator. Lower panel: corresponding schematic depictions of
the electronic structure and JT molecular distortions for the
metallic state and MJT insulator. Both in the metallic state and
in the MI state the molecules are distorted due to el.-vib.
interaction. As the lattice expands the distortions enlarge. If
the localization radii of electrons on neighboring molecules
overlap, then the electrons are collectivized and the metallic
state takes place (four molecules are pictured for example). If
the localization radius is smaller than intermolecular distance,
then electrons can be localized by the Coulomb correlations and
the MI state occurs.} \label{Fig7}
\end{figure}

Following to \cite{nom4}, the change of the potential $\Delta
V_{\texttt{SCF}}$, on which electrons are scattered, is given by a
sum of the change in the ionic potential $\Delta V_{\texttt{ion}}$
and the screening contribution from the Hartree and exchange
channels, that can be reduced to $\Delta V_{\texttt{SCF}}=\Delta
V_{\texttt{ion}}/\varepsilon$. Since the electron-phonon coupling
$\lambda$ represents the scattering of electrons by $\Delta
V_{\texttt{SCF}}$, the screening process for el.-vib. coupling can
be decomposed in the very same way as that of $\Delta
V_{\texttt{SCF}}$:
\begin{equation}\label{1.24a}
    \lambda_{\texttt{cRPA}}=\frac{\lambda_{\texttt{bare}}}{\varepsilon_{\texttt{cRPA}}}.
\end{equation}
In the same time the experimentally observed vibron frequencies in
$\texttt{K}_{3}\texttt{C}_{60}$ \cite{zhou} differ little from the
vibron frequencies in $\texttt{C}_{60}$. Calculations in
\cite{nom4} show that the screening has weak effect on the
frequencies. Apparently, the oscillations of a fullerene molecule
are determined with internal elastic constants and little depend
on the external environment (the relationship
$\omega_{\texttt{RPA}}=\omega_{\texttt{bare}}/\varepsilon$ is
correct in the "jelly" model for metals only). Moreover, in the
case of alkali-doped fullerides, the el.-vib. coupling of the
individual mode is not large, while the accumulation of their
contributions leads to the total el.-vib. coupling of $g\sim
0.5\ldots 1$. Therefore, we do not expect a large difference
between bare $\omega_{0}$ and the screened frequencies. Since
$\widetilde{U}_{vib}\propto\frac{2\lambda^{2}}{\omega_{0}}$ then
\begin{equation}\label{1.24}
    \widetilde{U}_{vib}^{\texttt{cRPA}}\simeq\frac{\widetilde{U}_{vib}^{\texttt{bare}}}{\varepsilon^{2}_{\texttt{cRPA}}}.
\end{equation}
Thus we can see that as the crystal lattice expands (the bandwidth
$W$ decreases) the density of state in the conduction band
increases - Eq.(\ref{1.14}) and the el.-el. interaction via
vibrons increases too - Eq.(\ref{1.24}). In the same time the
Coulomb barrier $U-U'+3J$ for the pairing enlarges -
Eq.(\ref{1.23}), Tab.(\ref{tab1}), but slower than
$\widetilde{U}_{vib}^{\texttt{cRPA}}$. Thus as volume per
$\texttt{C}_{60}$ enlarges then $T_{c}$ increases. However, in the
same time, the Coulomb correlations are amplified (the bandwidth
$W$ narrows and the on-site Coulomb repulsion grows), hence the
renormalization parameter $Z$ decreases - Eq.(\ref{1.20}). This
suppresses the critical temperature - Eq.(\ref{1.22}). At some
critical value of $W$, where $Z=0$, transition to MI state occurs
and $T_{c}=0$ in this point. Thus we have a \emph{dome} shaped
line $T_{c}$ which is shown in Fig.\ref{Fig7} as the line (a): for
weakly correlated regime ($Z\rightarrow 1$) $T_{c}$ increases with
decreasing of $W$ (with enlarging of volume per molecule
$C_{60}$), for strongly correlated regime ($Z\rightarrow 0$)
$T_{c}$ decreases with decreasing of $W$.

We could see that the effectiveness of el.-vib. interaction is
decreasing as temperature rises - Eqs.(\ref{1.3e5},\ref{1.3e6}).
So for small $T$ we have
$U_{vib}\propto\frac{2\lambda^{2}}{\omega_{0}}\left(1-\frac{T}{\omega_{0}}\right)$.
Thus the function $\Delta E=U-V-\widetilde{U}_{vib}(T)$ is
increasing with temperature, then at some temperature
$T_{\texttt{MI}}$ the criterion (\ref{1.7b}) of transition from
the metallic state to the Mott insulator state will be satisfied:
\begin{equation}\label{1.25}
\frac{\Delta
E}{W}=\frac{U-V-\widetilde{U}_{vib}(T_{\texttt{MI}})}{W}=1.
\end{equation}
We can see that at decreasing of $W$ the temperature
$T_{\texttt{MI}}$ drops. This dependence is schematically shown in
Fig.\ref{Fig7} as the line (b). It should be noted that at nonzero
temperatures the transition between metallic and insulator phases
is blurred because temperature is larger than the energetic gap
between Hubbard subbands. Hence the line separating metal and Mott
insulator phases at nonzero temperature $T_{\texttt{MI}}$ is more
crossover than phase transition.

\subsection{Influence of Jahn-Teller deformations}\label{polaron}

According to JT theorem any non-linear molecular system in a
degenerate electronic state will be unstable and will undergo
distortion to form a system of lower symmetry and lower energy
thereby removing the degeneracy. Thus a fullerene molecule with
partially filled $t_{1u}$ orbital ($\texttt{C}_{60}^{n-}$,
$n=1\ldots 5$) must be distorted (to $D_{2h}$ symmetry) due to the
el.-vib. interaction and the degeneracy of electronic state must
be removed. In the same time the molecules are combined into a
crystal and their electrons are collectivized. The collectivized
electrons interact with the molecular distortions and formation of
polarons (bound state of an electron with induced deformation by
it) can take place. According to Eq.(\ref{1.16b}) the binding
energy is $\sim\frac{\lambda^{2}}{\omega_{0}}\sim U_{vib}$. Then
according to the uncertainty principle the relation
$U_{vib}\frac{l}{v_{F}}\sim\hbar$ occurs, where $l$ is
localization radius of an electron, $v_{F}$ is Fermi velocity.
Following to \cite{dzum4} the electron can be localized in the
induced deformation if the localization radius is less than
average distance between electrons $\langle r\rangle$, which
determines Fermi energy
$\varepsilon_{F}=\frac{mv_{F}^{2}}{2}=\frac{\hbar^{2}}{2m}\left(\frac{3\pi^{2}N}{V}\right)^{2/3}\Rightarrow\langle
r\rangle\sim\sqrt[3]{V/N}\sim \hbar/\sqrt{\varepsilon_{F}m}$. Then
the condition $l=\langle r\rangle$ is
\begin{equation}\label{1.26a}
    U_{vib}\sim\varepsilon_{F}=\frac{W}{2}.
\end{equation}
Thus we have lattice of JT distorted molecules with collectivized
electrons if $U_{vib}\ll W$. As lattice expands the distortions of
the molecules increase but electrons cannot be localized because
the localization radius is larger than average distance between
electrons (and intermolecular distance). This state can be called
the Jahn-Teller metal which is observed in \cite{zadik}. When the
localization radius becomes equal and less than the average
distance between electrons, then formation of polaron of small
radius (the Holstein polaron) occurs \cite{alex2}. Such polaron
can be localized on a site by the Coulomb blockade $\sim U$ and we
obtain the Mott-Jahn-Teller insulator. Since the electrons form
the paired states on each molecule (the local pairs), which, in
turn, are locked on the molecules by the Coulomb correlations,
then the bosonization of Cooper pairs occurs. These configurations
(JT metal and MJT insulator) are schematically illustrated in the
lower panel of Fig.\ref{Fig7}.

The Hamiltonian (\ref{1.1},\ref{1.2}) is similar to the
Holstein-Hubbard Hamiltonian:
\begin{eqnarray}\label{1.26}
\widehat{H}&=&-t\sum_{\langle ij\rangle}\sum_{\sigma}
a^{+}_{i\sigma}a_{j\sigma}-\mu\sum_{i}\left(n_{i\uparrow}+n_{i\downarrow}\right)
+U\sum_{i}n_{i\uparrow}n_{i\downarrow}+\lambda\sum_{i}\left(n_{i\uparrow}+n_{i\downarrow}-1\right)\left(b_{i}+b^{+}_{i}\right)
+\omega_{0}\sum_{i}b^{+}_{i}b_{i},
\end{eqnarray}
Using Lang-Firsov canonical transformation
$e^{S}\widehat{H}e^{-S}$ \cite{wer1,wer2,alex1}, where
$S=\frac{\lambda}{\omega_{0}}\sum_{i}\left(n_{i\uparrow}+n_{i\downarrow}\right)\left(b_{i}^{+}-b_{i}\right)$,
and projection onto the subspace of zero phonons,
$H_{\texttt{LF}}=\left\langle
0\left|e^{S}\widehat{H}e^{-S}\right|0\right\rangle$, the
Hamiltonian (\ref{1.26}) is diagonalized:
\begin{eqnarray}\label{1.27}
\widehat{H}_{\texttt{LF}}&=&-te^{-\frac{\lambda^{2}}{\omega_{0}^{2}}}\sum_{\langle
ij\rangle}\sum_{\sigma}
a^{+}_{i\sigma}a_{j\sigma}-\mu_{eff}\sum_{i}\left(n_{i\uparrow}+n_{i\downarrow}\right)
+\left(U-\frac{2\lambda^{2}}{\omega_{0}}\right)\sum_{i}n_{i\uparrow}n_{i\downarrow},
\end{eqnarray}
where
\begin{equation}\label{1.28}
    \mu_{eff}=\mu-\frac{\lambda^{2}}{\omega_{0}}.
\end{equation}
Thus we obtain effective el.-el. interaction reduced to BCS-like
interaction (point, nonretarded):
$U-\frac{2\lambda^{2}}{\omega_{0}}$. The hopping is renormalized
as $t\rightarrow te^{-\frac{\lambda^{2}}{\omega_{0}^{2}}}$, thus
the collapse of the conduction band to the narrow polaron band
occurs. This polaron is a statical deformation of a molecule,
where an electron is in potential well of depth
$\sim\lambda^{2}/\omega_{0}$ and it supports the deformation by
own field. From Eq.(\ref{1.28}) we can see that the chemical
potential $\mu=W/2$ falls by $\frac{\lambda^{2}}{\omega_{0}}$. If
$\mu_{eff}<0$, then the Fermi level falls below the bottom of
conduction band, hence the localization of an electron on a site
by formation of local deformation (the Holstein polaron) becomes
possible. Then the condition of collapse of conduction band is
\begin{equation}\label{1.29}
    \frac{\lambda^{2}}{\omega_{0}}=\frac{W}{2}.
\end{equation}
This equation can be rewritten as $\frac{U_{vib}}{2}=\frac{W}{2}$,
where $U_{vib}=\frac{2\lambda^{2}}{\omega_{0}}$ is the attraction
energy in a pair, that corresponds to Eq.(\ref{1.26a}) obtained
from the uncertainty principle. When deriving
Egs.(\ref{1.15},\ref{1.16}), we have seen that the multi-band
system can be reduced to an effective single band superconductor.
Thus for the case of interaction with $A_{g}$ and $H_{g}$ phonons
the condition (\ref{1.29}) will have a form:
\begin{equation}\label{1.30}
    \frac{\widetilde{U}_{vib}}{2}=\frac{W}{2}\Rightarrow g=1.
\end{equation}

Formation of the JT deformation leads from the Hund's electron
configuration $|\uparrow,\uparrow,\uparrow\rangle$ to the local
pair configuration $|\uparrow\downarrow,\uparrow,0\rangle$. As we
have seen above in order to make the local pair on a site we have
to overcome the configurational Coulomb barrier $U-U'+3J$ -
Fig.(\ref{Fig4}). If the JT energy $\propto
\lambda^{2}/\omega_{0}^{2}$ is less than Hund's coupling $\propto
J$, that, as it has been shown in \cite{wehrli}, the JT effect is
completely suppressed. In formation of configuration with a local
pair three electrons take part, hence the energy $(U-U'+3J)/3$ is
per each electron. Since this energy resists the formation of the
polaron, then Eq.(\ref{1.30}) should be refined as
\begin{equation}\label{1.31}
    \frac{\widetilde{U}_{vib}}{2}-\frac{U-U'+3J}{3}=\frac{W}{2}\Rightarrow g-\frac{2}{3}\mu_{c}=1.
\end{equation}
The polaron narrowing of the conduction band enhances Coulomb
correlations in the already highly correlated system:
$We^{-\frac{\lambda^{2}}{\omega_{0}^{2}}}\ll W<U$, that must turn
the material to the Mott insulator. Above we could see that as the
crystal lattice expands (volume per $\texttt{C}_{60}$ increases)
the bandwidth $W$ decreases and the el.-vib. coupling
$\widetilde{U}_{vib}$ enlarges. Thus increasing the volume per the
molecule we reach the border (\ref{1.31}). This border is shown in
Fig.\ref{Fig7} as the line (c). Then the conduction band collapses
and the material becomes Mott-Jahn-Teller insulator. Since a Mott
insulator is antiferromagnetic at half-filled conduction band
\cite{penn,izyum}, hence the MJT-insulating phase of
$\texttt{A}_{3}\texttt{C}_{60}$ should be magnetically ordered at
low temperature.

\begin{table}\caption{Vibrational energies $A_{g}$ and $H_{g}$ of $\texttt{C}_{60}$ (from \cite{koch}).}\label{tab2}
\begin{center}
\begin{tabular}{|c|cccccccccc|}
  \hline\rule{0cm}{0.5cm}
    $\omega_{M}$ &$A_{g}(1)$ &$A_{g}(2)$&$H_{g}(1)$&$H_{g}(2)$&$H_{g}(3)$&$H_{g}(4)$&$H_{g}(5)$&$H_{g}(6)$&$H_{g}(7)$&$H_{g}(8)$ \\[0.1cm]
  \hline\rule{0cm}{0.5cm}
    $\texttt{K}$ & 717 & 2117 & 393 & 624 & 1024 & 1116 & 1585 & 1801 & 2053 & 2271 \\[0.1cm]
\hline\rule{0cm}{0.5cm}
    $\texttt{meV}$ & 62 & 182 & 34 & 54 & 88 & 96 & 137 & 155 & 177 & 196 \\[0.1cm]
  \hline
\end{tabular}
\end{center}
\end{table}

\subsection{Calculation of the phase diagram}\label{phase}

\begin{table}\caption{ Calculated characteristics of $\texttt{A}_{3}\texttt{C}_{60}$ (where $\texttt{A=K,Rb,Cs}$
at normal pressure and $\texttt{A=Cs}$ at pressures
7\texttt{kbar}, 2\texttt{kbar}) as function of volume per
$\texttt{C}_{60}$ neglecting thermal expansion of the lattice: the
Mott parameters corresponding to Hund's rule without el.-vib.
interaction $\frac{U+4J-V}{W}$ and to presence of the local pairs
on sites $\frac{U-V-\widetilde{U}_{vib}}{W}$ (anti-Hund's rule),
renormalization parameter $Z$, the el.-vib. coupling constant
$\lambda$ (for $\texttt{K}_{3}\texttt{C}_{60}$ the constant is an
adjustable parameter - bold font), energy of attraction in the
local pair $\widetilde{U}_{vib}$ at $T=0$, el.-el. attraction
constant $g$ at $T=0$, el.-el. repulsion constant $\mu_{c}$,the
parameter $g-\frac{2}{3}\mu_{c}$ (at the border of collapse of
conduction band it is equal to $1$), the critical temperature
$T_{c}$ (in brackets the values of $T_{c}$ are given which would
be if $\Omega=\omega_{0}$ is supposed), the temperature of the
Mott transition $T_{\texttt{\texttt{MI}}}$ (for
$\texttt{K}_{3}\texttt{C}_{60}$ this temperature is so high that
makes no sense). For $\texttt{Cs}_{3}\texttt{C}_{60}$ at normal
pressure the parameters, which would be if the conduction band
does not collapse, are shown in italic.}\label{tab3}
\begin{center}\label{tab3}
\begin{tabular}{|c|ccccc|}
  \hline\rule{0cm}{0.5cm}
    & $\texttt{K}_{3}\texttt{C}_{60}$ & $\texttt{Rb}_{3}\texttt{C}_{60}$ & $\texttt{Cs}_{3}\texttt{C}_{60}$ (7\texttt{kbar}) & $\texttt{Cs}_{3}\texttt{C}_{60}$ (2\texttt{kbar}) & $\texttt{Cs}_{3}\texttt{C}_{60}$  \\[0.1cm]
  \hline\rule{0cm}{0.5cm}
  volume per $\texttt{C}_{60}$ $ (\texttt{A}^{3})$ & 722 & 750 & 762 & 784 & 804 \\
  \hline\rule{0cm}{0.5cm}
  $\frac{\Delta E}{W}=\frac{U+4J-V}{W}$ & 1.39 & 1.74 & 1.89 & 2.31 & \emph{2.68} \\[0.1cm]
  \rule{0cm}{0.5cm}
  experimental $T_{c}(\texttt{K})$ & 19 & 29 & 35 & 26 & - \\[0.1cm]
\rule{0cm}{0.5cm}
  $\lambda(\texttt{meV})$ & \textbf{38.22} & 41.97 & 43.68 & 46.52 &  48.64\\[0.1cm]
  \rule{0cm}{0.5cm}
  $\widetilde{U}_{vib}(\texttt{eV})$ & 0.34 & 0.41 & 0.44 & 0.50 &  0.55\\[0.1cm]
\rule{0cm}{0.5cm}
  $g$ & 0.67 & 0.90 & 1.04 & 1.32 &  \emph{1.61}\\[0.1cm]
\rule{0cm}{0.5cm}
  $\mu_{c}$ & 0.30 & 0.38 & 0.41 & 0.49 &  \textit{0.52}\\[0.1cm]
\rule{0cm}{0.5cm}
  $g-\frac{2}{3}\mu_{c}$ & 0.47 & 0.65 & 0.76 & 1.00 &  \emph{1.26}\\[0.1cm]
  \rule{0cm}{0.5cm}
  $\frac{\Delta E}{W}=\frac{U-V-\widetilde{U}_{vib}}{W}$ & 0.47 & 0.54 & 0.52 & 0.61 &  \emph{0.65}\\[0.1cm]
\rule{0cm}{0.4cm}
$Z$ & 0.78 & 0.71 & 0.73 & 0.62 &  \emph{0.58}\\[0.1cm]
\rule{0cm}{0.4cm}
 calculated $T_{c}(\texttt{K})$ & 19 & 26 & 43 & 32 (33) & \emph{31} (\emph{33}) \\[0.1cm]
  \rule{0cm}{0.5cm}
  $T_{\texttt{\texttt{MI}}}(\texttt{K})$ & - & 450 & 400 & 215 & \emph{155} \\[0.1cm]
\hline
\end{tabular}
\end{center}
\end{table}
\begin{figure}[h]
\includegraphics[width=10cm]{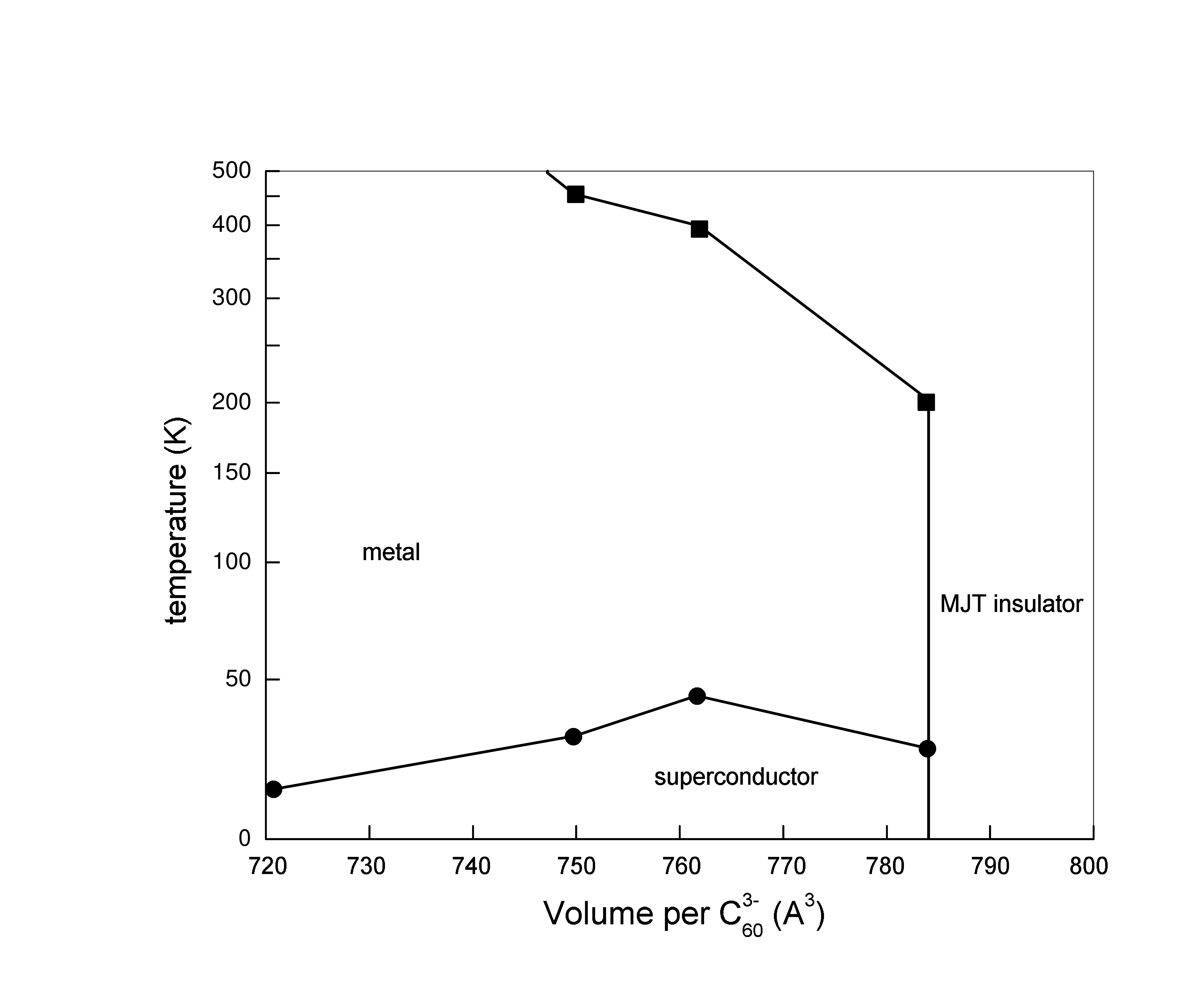}
\caption{Results of calculations for $T_{c}$ (circle markers),
$T_{\texttt{MI}}$ (square markers) and the border of collapse of
conduction band (vertical line) from Tab.\ref{tab3}. Thus we have
a phase diagram of of alkali-doped fullerides as a function of
volume per $\texttt{C}_{60}$ and temperature.} \label{Fig8}
\end{figure}
\begin{figure}[h]
\includegraphics[width=16cm]{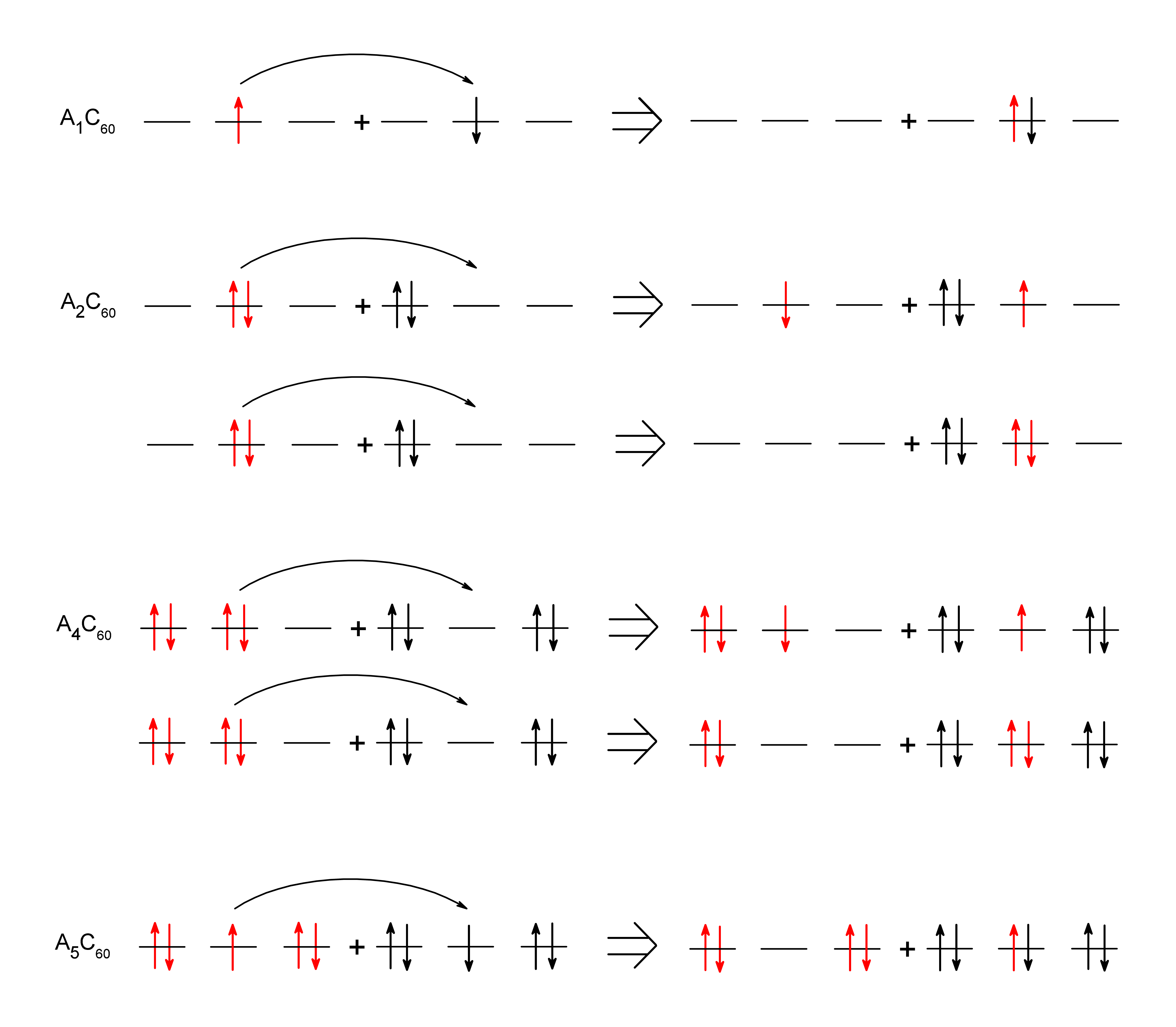}
\caption{The charge transfer in materials
$\texttt{A}_{n}\texttt{C}_{60}$ ($n=1,2,4,5$)} \label{Fig9}
\end{figure}

Using obtained results (\ref{1.22},\ref{1.25},\ref{1.31}), values
of bandwidth $W$ and Coulomb parameters $U$, $U'$, $J$, $V$ from
Tab.\ref{tab1} we can calculate $T_{c}$, $T_{\texttt{MI}}$, border
of the collapse of conduction band (\ref{1.31}) and the Mott
parameter $\frac{\Delta E}{W}$ with $\Delta E$ from
Eqs.(\ref{1.7a},\ref{1.7d}) for $\texttt{A}_{3}\texttt{C}_{60}$
(where $\texttt{A}=\texttt{K},\texttt{Rb},\texttt{Cs}$ and the
substance with cesium is considered at normal pressure,
$2\texttt{kbar}$ and $7\texttt{kbar}$ with $\texttt{fcc}$
structure). However we must know parameters $\lambda_{\nu}$ it
order to calculate the el.-el. coupling via vibrons
$\widetilde{U}_{vib}$ - Eqs.(\ref{1.3e5},\ref{1.3e6}). We can use
an adjustable parameter: for $\texttt{K}_{3}\texttt{C}_{60}$ the
critical temperature is $T_{c}=19\texttt{K}$, then we choose the
parameter of el.-vib. coupling $\lambda$ so that the critical
temperature calculated from (\ref{1.22}) would be equal to the
experimental one. For other materials we calculate the parameter
$\lambda$ using Eqs.(\ref{1.24a},\ref{1.24}), for example
\begin{equation}\label{1.32}
    \lambda(\texttt{Rb})=\lambda(\texttt{K})
\frac{\varepsilon(\texttt{Rb})}{\varepsilon(\texttt{K})}
\Rightarrow\widetilde{U}_{vib}(\texttt{Rb})=\widetilde{U}_{vib}(\texttt{K})
\frac{\varepsilon^{2}(\texttt{Rb})}{\varepsilon^{2}(\texttt{K})}.
\end{equation}
Moreover we must known the vibron frequency $\omega_{0}$. In a
fullerene molecule an electron interacts with $A_{g}$ and $H_{g}$
vibrational modes presented in Tab.\ref{tab2}. Then the parameter
$\widetilde{U}_{vib}$ can be calculated with a following manner:
\begin{equation}\label{1.32}
    \widetilde{U}_{vib}=2\lambda^{2}\sum_{M=A_{g},H_{g}}\frac{1}{\omega_{M}}\left[\coth\left(\frac{\omega_{M}}{T}\right)-\frac{T}{\omega_{M}}\right]
\end{equation}
Here the coupling constants $\lambda_{\nu}$ has been replaced by
effective coupling constant $\lambda$:
$\sum_{\nu}\lambda^{2}_{\nu}\left[V^{(\nu)}_{mm}V^{(\nu)}_{mm}+\sum_{m'\neq
m}V^{(\nu)}_{mm'}V^{(\nu)}_{mm'}\right]\rightarrow \lambda^{2}$,
because we use $\lambda$ as an adjustable parameter. The integral
(\ref{1.22}) can be cut off by the largest vibron energy
$\omega_{0}=\omega(H_{g}(8))=0.196\texttt{eV}$, that is confirmed
by numerical calculation for vibron-mediated interactions in
$\texttt{Cs}_{3}\texttt{C}_{60}$ in \cite{nom4} where
$\omega_{0}\approx 0.19\texttt{eV}$, or it can be cut off by the
half-bandwidth $W/2$ if $2\omega_{0}>W$. Results of the
calculations, \emph{neglecting thermal expansion of the lattice},
are presented in Tab.(\ref{tab3}) and Fig.\ref{Fig8}. From the
table we can see that without the el.-vib. interaction we have
$\frac{\Delta E}{W}>1$, hence all materials would be Mott
insulators. However the el.-vib. interaction and the local pairing
change the relation as $\frac{\Delta E}{W}<1$, hence these
materials becomes conductors, in the same time the Coulomb
correlations enlarges as the lattice constant increases. The
parameter $g-\frac{2}{3}\mu_{c}$ at volume per $\texttt{C}_{60}$
$784\texttt{A}^{3}$ becomes $1$, hence material
$\texttt{Cs}_{3}\texttt{C}_{60}$ at pressure $2\texttt{kbar}$ is
near the border of collapse of conduction band, and at normal
pressure the material is Mott-Jahn-Teller insulator. As noted
above at the border of collapse the localization radius becomes
equal to the average distance between electrons, thus formation of
polaron of small radius occurs. Such polaron can be localized on a
site by the Coulomb blockade $U$. It should be noted that since
$\omega_{0}\approx W/2$ for $\texttt{Cs}_{3}\texttt{C}_{60}$ at
pressure $2\texttt{kbar}$ and normal pressure, then the cutting
off the integral (\ref{1.22}) by $W/2$ instead by $\omega_{0}$
does not influence on the results significantly. From
Fig.\ref{Fig8} we can see that the calculated phase diagram of
alkali-doped fullirides is quantitatively close to experimental
phase diagram in Fig.\ref{Fig1}.

\section{Conductivity of $\texttt{A}_{n}\texttt{C}_{60}$ with
$n=1,2,4,5$}\label{family}

For materials $\texttt{A}_{n}\texttt{C}_{60}$, where $n=1,2,4,5$,
we can suppose that the el.-el. interaction $\widetilde{U}_{vib}$
via vibrons is approximately the same for these materials and is
equal to the value in $\texttt{A}_{3}\texttt{C}_{60}$ $\sim
0.4\texttt{eV}$, it is analogously for Coulomb $U,U',V$ and
exchange $J$ interactions. The charge transfer in these materials
is shown in Fig.\ref{Fig9}.

\begin{itemize}
    \item $\texttt{A}_{1}\texttt{C}_{60}$. In order to form a local pair we have to transfer an electron from a site
    to a nearby site containing another electron. For this it is
    necessary to make a positive work $U-V-\widetilde{U}_{vib}>0$. Thus formation of the pairs is energetically unfavorable,
    i.e. the pairs are unstable.  In the same time $\frac{\Delta E}{W}=\frac{U-V-\widetilde{U}_{vib}}{W}<1$,
    because we create the pair on a neighboring molecule.
    Thus this material is a conductor due the el.-vib. interaction but it is not a superconductor.
    \item $\texttt{A}_{2}\texttt{C}_{60}$. Two electrons are in the paired state on a site because the energy
    of the state is $U-U'+J-\widetilde{U}_{vib}<0$ (measured from the Hund's rule state). In order to transfer an electron from a site to a site we must \emph{break} a pair. In this
    case $\frac{\Delta E}{W}=\frac{2U'-U-V+\widetilde{U}_{vib}}{W}>1$. Thus the transfer of a electron is blocked by Coulomb interaction and
    the el.-el. attraction via vibrons $\widetilde{U}_{vib}$. Then the pair is compact and it could be transferred but
    $\frac{\Delta E}{W}=\frac{2U'-4V}{W}>1$. Hence this material is insulator and a molecule $\texttt{C}_{60}^{2-}$ does not have spin.
    \item $\texttt{A}_{4}\texttt{C}_{60}$. Like the previous material all electrons are in the paired state on a site
    because the energy of this state is negative. We can transfer the charge from a site to a site by transferring of an electron with breaking of
    the pair or by transferring of the compact pair. In these cases $\frac{\Delta E}{W}$ is $\frac{2U'-U-V+\widetilde{U}_{vib}}{W}>1$
    and $\frac{2U'-4V}{W}>1$ accordingly. Thus this material is insulator and a molecule $\texttt{C}_{60}^{4-}$ does not have spin.
    \item $\texttt{A}_{5}\texttt{C}_{60}$. In order to transfer an electron from a site
    to a nearby site we have to spend such energy that $\frac{\Delta E}{W}=\frac{3U-2U'-V-\widetilde{U}_{vib}}{W}<1$.
    This process forms the pair, but $3U-2U'-V-\widetilde{U}_{vib}>0$ - energy of the pair is positive,
    hence the Cooper pairs are not stable like in $\texttt{A}_{1}\texttt{C}_{60}$.
    Thus this material is conductor due to the el.-vib. interaction but it is not superconductor.
\end{itemize}
As expected, since the system is particle-hole symmetric,
therefore the properties for $n=1,2$ are identically with the
properties for $n=5,4$ accordingly. Thus we can see the el.-vib.
interaction transforms Mott insulators
$\texttt{A}_{n}\texttt{C}_{60}$ with $n=1,3,5$ to conductors.
However for $n=2,4$ this interaction prevents conduction and it
pairs electrons on each molecules so that the molecules do not
have spin.

\section{Results}\label{concl}

We have considered the problem of conductivity and
superconductivity of alkali-doped fullerides
$\texttt{A}_{n}\texttt{C}_{60}$
($\texttt{A}=\texttt{K},\texttt{Rb},\texttt{Cs}$, $n=1\ldots 5$)
while these materials would have to be antiferromagnetic Mott
insulators because the on-site Coulomb repulsion is larger than
bandwidth: $U\sim 1\texttt{eV}>W\sim 0.5\texttt{eV}$, and
electrons on a molecule have to be distributed over molecular
orbitals according to the Hund's rule. We have found important
role of 3-fold degeneration of LUMO ($t_{1u}$ level), small
hopping between neighboring molecules and the coupling of
electrons to Jahn-Teller modes (vibrons). The el.-el. coupling via
vibrons $U_{vib}$ cannot compete with the on-site Coulomb
repulsion $U\gg U_{vib}$, but it can compete with the Hund
coupling $U-U'+3J\approx5J\sim U_{vib}$ (where the exchange energy
is much less than direct Coulomb interaction $J\ll U$). This
allows to form the local pair (\ref{1.4a}) on a molecule.
Formation of the local pairs radically changes conductivity of
these materials: they can make $\frac{\Delta E}{W}<1$, where
$\Delta E$ is the energy change at transfer of an electron from a
site to nearby site, while without the el.-vib. interaction we
have $\frac{\Delta E}{W}> 1$ that corresponds to Mott insulator.
We have shown that the el.-vib. interaction and the local pairing
transform Mott insulators $\texttt{A}_{n}\texttt{C}_{60}$ with
$n=1,3,5$ to conductors. However for $n=2,4$ this interaction
prevents conduction and it pairs electrons on each molecules so
that the molecules do not have spin. The local pair mechanism can
ensure superconductivity of $\texttt{A}_{3}\texttt{C}_{60}$ which
is result of interplay between the el.-el. coupling via vibrons,
the Coulomb blockade on a site and the hopping between neighboring
sites. It should be noted that the size of a pair is larger than
average distance between electrons (between molecules) and the
pair is smeared over the crystal. That is in the metallic phase
the local pairs have fermionic nature. At the border of transition
to Mott-Jahn-Teller insulator the bosonization of local pairs
occurs due to Coulomb blockage of hopping of electron between
neighboring molecules. Thus system with the local pairing can be
effectively described by BCS theory. In such system we have the
effective coupling constant as $g-\mu_{c}>0$ (where $g$ is
determined with the el.-vib. interaction and $\mu_{c}$ is
determined with the Hund coupling - Eq.(\ref{1.16})) unlike usual
metal superconductors where $g-\mu^{*}_{c}>0$ only (where
$\mu^{*}_{c}$ is a Coulomb pseudopotential with Tolmachev's
reduction).

Since $\texttt{A}_{3}\texttt{C}_{60}$ is a strongly correlated
system, i.e. $U/W\sim 1$, then the equation for the critical
temperature (\ref{1.22}), unlike ordinary BCS equation, accounts
inter-site Coulomb correlations by renormalization parameter
(\ref{1.20}), which is $Z=0$ at transition to Mott insulator. The
correlations are amplified when the conduction band compressions
and they significantly suppress the critical temperature $T_{c}$.
In the same time, as the crystal lattice expands (the bandwidth
$W$ decreases) the density of states in the conduction band
increases as $1/W$ and the el.-vib. interaction intensifies due to
weakening of the screening. As a result the depending of $T_{c}$
on the volume per a molecule has a dome shape -
Figs.(\ref{Fig7},\ref{Fig8}), unlike the theoretical phase diagram
calculated by DMFT method in \cite{nom1,nom2}. Thus el.-vib
interaction ensures conductivity and superconductivity of
alkali-doped fullerides. However the effectiveness of el.-vib.
interaction is decreasing as temperature rises due to a vibron
propagator $\frac{2\omega_{0}}{(\pi nT)^{2}+\omega_{0}^{2}}$.
Hence at some temperature $T_{\texttt{MI}}$ the criterion of
transition from the metallic state to the Mott insulator state
will be satisfied - Eq.(\ref{1.25}), and the material becomes Mott
insulator.

As the bandwidth decreases the collapse of the conduction band to
the narrow polaron band occurs when the condition (\ref{1.31}) is
satisfied. This Holstein polaron is a statical JT deformation of
the molecule, where an electron supports the deformation by own
field. According to JT theorem a fullerene molecule with partially
filled $t_{1u}$ orbital must be distorted due to the el.-vib.
interaction and the degeneracy of electron state must be removed.
In the same time the molecules are combined into a lattice and
their electrons are collectivized. As the lattice expands the
distortions of the molecules increase, but electrons cannot be
localized because the localization radius is larger than average
distance between electrons (and intermolecular distance), i.e. the
electrons are collectivized. This state can be called the
Jahn-Teller metal which is observed in \cite{zadik}. When the
localization radius becomes equal and less than the average
distance between electrons, then formation of polaron of small
radius occurs. The polaron narrowing of the conduction band
enhances Coulomb correlations in already strongly correlated
system, that must turn the alkali-doped fullerides to the Mott
insulator, and the material becomes Mott-Jahn-Teller insulator. We
have demonstrated that border of the collapse of conduction band
vertically cuts off the SC and metallic phases (at low
temperature) - Figs.(\ref{Fig7},\ref{Fig8}), unlike the
theoretical phase diagram calculated in \cite{hosh}.

Thus we have three phases of $\texttt{A}_{3}\texttt{C}_{60}$
illustrated schematically in Fig.(\ref{Fig7}): superconductor,
metal and MJT insulator, which are separated by lines $T_{c}$,
$T_{\texttt{MI}}$ and the border of collapse of conduction band.
Calculated phase diagram in Fig.\ref{Fig8} of alkali-doped
fullerides is quantitatively close to experimental phase diagram
in Fig.\ref{Fig1}. We have illustrated that
$\texttt{K}_{3}\texttt{C}_{60},\texttt{Rb}_{3}\texttt{C}_{60}$ are
conductors (and superconductors) but
$\texttt{Cs}_{3}\texttt{C}_{60}$ is MJT insulator at normal
pressure, at $2\texttt{kbar}$ it is superconductor on the border
with MJT insulator, and at $7\texttt{kbar}$ it is superconductor
with maximal $T_{c}$. Thus we have shown that superconductivity,
conductivity and insulation of alkali-doped fullerides have common
nature: the local pairing due to interaction with the Jahn-Teller
phonons. The proposed model does not account influence of the
crystal field, therefore we consider materials only with the
merohedrally disordered $\texttt{fcc}$ structure unlike the
ordered $\texttt{A}15$ structure where the effect of crystal field
should be stronger.

\acknowledgments

This research was supported by theme grant of department of
physics and astronomy of NAS of Ukraine: "Dynamics of formation of
spatially non-uniform structures in many-body systems", PK
0118U003535



\begin{thebibliography}{99}

\bibitem{zadik} R. H. Zadik, Y. Takabayashi, G. Klupp, R. H. Colman, A. Y. Ganin, A. Potoиnik, P. Jegliи,
D. Arиon, P. Matus, K. Kamarбs, Y. Kasahara, Y. Iwasa, A. N.
Fitch, Y. Ohishi, G. Garbarino, K. Kato, M. J. Rosseinsky, K.
Prassides , Science Advances  \textbf{1}, no.3, e1500059 (2015),
https://doi.org/10.1126/sciadv.1500059

\bibitem{klupp} Gyongyi Klupp, Peter Matus, Katalin Kamaras, Alexey Y. Ganin, Alec McLennan, Matthew J. Rosseinsky,
Yasuhiro Takabayashi, Martin T. McDonald, Kosmas Prassides, Nature
communications  \textbf{3}, 912 (2012),
https://doi.org/10.1038/ncomms1910

\bibitem{kamar} Katalin Kamaras, Gyongyi Klupp, Peter Matus, Alexey Y Ganin, Alec McLennan, Matthew J Rosseinsky,
Yasuhiro Takabayashi, Martin T McDonald, Kosmas Prassides, Journal
of Physics: Conference Series \textbf{428}, 012002 (2013),
https://doi.org/10.1088/1742-6596/428/1/012002

\bibitem{bald} L. Baldassarre, A. Perucchi, M. Mitrano, D. Nicoletti, C. Marini, D. Pontiroli,
M. Mazzani, M. Aramini, M. Ricc, G. Giovannetti, M. Capone, S.
Lupi, Scientific Reports \textbf{5}, 15240 (2015),
https://doi.org/10.1038/srep15240

\bibitem{takab} Yasuhiro Takabayashi, Kosmas Prassides, Phil. Trans. R. Soc. A \textbf{374}, 20150320
(2016), http://dx.doi.org/10.1098/rsta.2015.0320


\bibitem{alloul1} V. Brouet, H. Alloul, Thien-Nga Le, S. Garaj, L. Forrу, Phys. Rev. Lett.
\textbf{86}, 4680 (2001),
https://doi.org/10.1103/PhysRevLett.86.4680

\bibitem{alloul2} V. Brouet, H. Alloul, S. Garaj, L. Forro, Phys. Rev. B \textbf{66}
155124 (2002), https://doi.org/10.1103/PhysRevB.66.155124

\bibitem{alloul3} Y. Ihara, H. Alloul, P. Wzietek, D. Pontiroli, M. Mazzani, M. Ricco, Phys. Rev. Lett.
\textbf{104}, 256402 (2010),
https://doi.org/10.1103/PhysRevLett.104.256402

\bibitem{alloul4} Y. Ihara, H. Alloul, P. Wzietek, D. Pontiroli, M. Mazzani, M. Ricco, EPL
\textbf{94},
37007 (2011), https://doi.org/10.1209/0295-5075/94/37007

\bibitem{alloul5} H. Alloul, P. Wzietek, T. Mito, D. Pontiroli, M. Aramini, M. Ricco, J.P. Itie, E. Elkaim, Phys. Rev. Lett.
\textbf{118},
237601 (2017), https://doi.org/10.1103/PhysRevLett.118.237601

\bibitem{gun0} O. Gunnarsson, Rev. Mod. Phys. \textbf{69}, 575
(1997), https://doi.org/10.1103/RevModPhys.69.575

\bibitem{chen1} Guanhua Chen, William A. Goddard III, Proc. Natl. Acad. Sci. USA
\textbf{90}, 1350 (1993), https://doi.org/10.1073/pnas.90.4.1350


\bibitem{cap} E. Cappelluti, C. Grimaldi, L. Pietronero, S. Str\"assler, G.A. Ummarino, Eur. Phys. J. B \textbf{21}
383 (2001), https://doi.org/10.1007/PL00011123


\bibitem{han1} J.E. Han, O. Gunnarsson, Physica B \textbf{292}, 196
(2000), https://doi.org/10.1016/S0921-4526(00)00482-8

\bibitem{han2} J.E. Han, O. Gunnarsson, V.H. Crespi, Phys. Rev. Lett. \textbf{90}, 167006
(2003), https://doi.org/10.1103/PhysRevLett.90.167006

\bibitem{lam1} Paul E. Lammert and Daniel S. Rokhsar, Phys. Rev. B \textbf{48}, 4103
(1993), https://doi.org/10.1103/PhysRevB.48.4103

\bibitem{lam2} D. M. Deaven, P. E. Lammert, D. S. Rokhsar, Phys. Rev. B \textbf{52}, 16377
(1995), https://doi.org/10.1103/PhysRevB.52.16377

\bibitem{koga} Akihisa Koga and Philipp Werner, Phys.Rev.B \textbf{91}, 085108
(2015), https://doi.org/10.1103/PhysRevB.91.085108

\bibitem{suz} Shugo Suzuki, Kenji Nakao, Phys. Rev. B \textbf{52}, 14206
(1995), https://doi.org/10.1103/PhysRevB.52.14206

\bibitem{gun1} Olle Gunnarsson, Erik Koch, Richard M. Martin, Phys. Rev. B \textbf{54}, R11026
(1996), https://doi.org/10.1103/PhysRevB.54.R11026

\bibitem{koch} Erik Koch, \textit{The doped Fullerenes.
A family of strongly correlated systems}, Max-Planck-Institut
f\"ur Festk\"orperforschung, Stuttgart, 2003

\bibitem{nom1} Yusuke Nomura, Shiro Sakai, Massimo Capone, Ryotaro Arita, Science Advances  \textbf{1}, e1500568 (2015), https://doi.org/10.1126/sciadv.1500568

\bibitem{nom2} Yusuke Nomura, Shiro Sakai, Massimo Capone, Ryotaro Arita, J. Phys.: Condens. Matter \textbf{28}, 153001 (2016), https://doi.org/10.1088/0953-8984/28/15/153001

\bibitem{hosh} Shintaro Hoshino, Philipp Werner, Phys. Rev. Lett. \textbf{118}, 177002 (2017), https://doi.org/10.1103/PhysRevLett.118.177002




\bibitem{sav} Michael R. Savina, Lawrence L. Lohr, Anthony H. Francis, Chem. Phys. Lett. \textbf{205}, 200
(1993), https://doi.org/10.1016/0009-2614(93)89230-F

\bibitem{allon} F. Allonso-Marroquin, J. Giraldo, A. Calles, J.J. Castro, Rev. Mex. Fis. S \textbf{44(3)}, 18
(1998) 

\bibitem{hadd} R.C. Haddon, Acc. Chem. Res. \textbf{25}, 127
(1992), https://doi.org/10.1021/ar00015a005


\bibitem{geor} Antoine Georges, Luca de’ Medici, Jernej Mravlje, Annual Reviews of Condensed Matter Physics \textbf{4}, 137
(2013), https://doi.org/10.1146/annurev-conmatphys-020911-125045

\bibitem{feng} Qingguo Feng, P.M. Oppeneer, Phys. Rev. B \textbf{86}, 035107 (2012), https://doi.org/10.1103/PhysRevB.86.035107

\bibitem{nom3} Nomura Y., Nakamura K., Arita R., Phys. Rev. B
\textbf{85},
155452 (2012), https://doi.org/10.1103/PhysRevB.85.155452

\bibitem{gun} Olle Gunnarsson, \textit{Alkali-Doped Fullerides
Narrow-Band Solids with Unusual Properties}, World Scientific
Publishing Co. Pte. Ltd., 2004,
https://doi.org/10.1142/9789812794956

\bibitem{tak} Hori C., Takada Y., \emph{Polarons and Bipolarons in Jahn–Teller Crystals},
Springer Series in Chemical Physics, vol 97., pp.841-871,
Springer, Berlin, Heidelberg (2009),
https://doi.org/10.1007/978-3-642-03432-9




\bibitem{ginz} V.L. Ginzburg, D.A. Kirzhnits, High-temperature superconductivity,
Consultants Bureau, New York and London, 1982

\bibitem{nom4} Yusuke Nomura, Ryotaro Arita, Phys. Rev. B \textbf{92},  245108 (2015), https://doi.org/10.1103/PhysRevB.92.245108

\bibitem{gebh}  Florian Gebhard, The Mott Metal-Insulator Transition. Models and Methods, Springer-Verlag Berlin Heidelberg, 1997, https://doi.org/10.1007/3-540-14858-2


\bibitem{rod} J.J. Rodr\'iguez-N\'u\~nez, A.A.Schmidt, Physica C: Superconductivity \textbf{350}, 88 (2001),
https://doi.org/10.1016/S0921-4534(00)01544-6

\bibitem{dzum1} S. Dzhumanov, E.X. Karimboev, Sh.S. Djumanov, Phys. Lett. A \textbf{380}, 2173 (2016), https://doi.org/10.1016/j.physleta.2016.04.038

\bibitem{litak} G. Litak, T. {\"O}rd, K. R\"ago and A. Vargunin, Acta Physica Polonica A \textbf{121}, 747
(2012), https://doi.org/10.12693/APhysPolA.121.747



\bibitem{asker1} I.N. Askerzade, Physica C \textbf{397}, 99 (2003), https://doi.org/10.1016/j.physc.2003.07.003

\bibitem{asker2} I.N. Askerzade, Physics-Uspekhi \textbf{49}, 1003
(2006), https://doi.org/10.1070/PU2006v049n10ABEH006055

\bibitem{asker3} I.N. Askerzade, J. Phys. Chem. Solids \textbf{68}, 1311 (2007), https://doi.org/10.1016/j.jpcs.2007.02.016

\bibitem{grig2} Konstantin V. Grigorishin, Phys. Lett. A \textbf{380}, 1781
(2016), https://doi.org/10.1016/j.physleta.2016.03.023

\bibitem{lok} V.M. Loktev, S.G. Sharapov, Cond. Matter Physics (Lviv) No.11., 131
(1997), http://dx.doi.org/10.5488/CMP.11.131

\bibitem{jones}  William Jones, Norman H. March, Theoreoretical solid state
physics, Vol.2, Dover Publications, Inc., New York, 1985.

\bibitem{brink} W.F. Brinkman, T.M. Rice, Phys. Rev. B \textbf{2}, 4302
(1970), https://doi.org/10.1103/PhysRevB.2.4302

\bibitem{bulla} R. Bulla, T.A. Costi, Vollhardt D., Phys.Rev.B \textbf{64}, 045103
(2001), https://doi.org/10.1103/PhysRevB.64.045103

\bibitem{gold} M.S. Golden, M. Knupfer, J. Fink, J.F. Armbruster, T.R.
Cummins, H.A. Romberg, M. Roth, M. Sing, M. Schmidt, E. Sohmen, J.
Phys.: Cond. Matter \textbf{7}, 8219 (1995),
https://doi.org/10.1088/0953-8984/7/43/004

\bibitem{zhou} Ping Zhou, Kai-An Wang, A.M. Rao, P.C. Eklund, G. Dresselhaus, M.S. Dresselhaus, Phys. Rev. B \textbf{45}, 10838(R)
(1992), https://doi.org/10.1103/PhysRevB.45.10838

\bibitem{dzum4} S. Dzhumanov, U.T. Kurbanov, Z.S. Khudayberdiev, A. R. Hafizov,  Low Temp. Phys \textbf{42}(11), 1057 (2016), http://dx.doi.org/10.1063/1.4971169

\bibitem{alex2} A.S. Alexandrov, A.B. Krebs, Sov. Phys. Usp. \textbf{35}, 345 (1992), https://doi.org/10.1070/PU1992v035n05ABEH002235

\bibitem{wer1} M. Casula, Ph. Werner, L. Vaugier, F. Aryasetiawan, T. Miyake, A. J. Millis, S. Biermann, Phys. Rev. Lett. \textbf{109}, 126408 (2012), https://doi.org/10.1103/PhysRevLett.109.126408

\bibitem{wer2} Yuta Murakami, Philipp Werner, Naoto Tsuji, Hideo Aok, Phys. Rev. B \textbf{88}, 125126 (2013), https://doi.org/10.1103/PhysRevB.88.125126

\bibitem{alex1} A.S. Alexandrov, Phys. Rev. B \textbf{61}, 12315 (2000), https://doi.org/10.1103/PhysRevB.61.12315

\bibitem{wehrli} S.Wehrli, M. Sigrist, Phys. Rev. B \textbf{76}, 125419 (2007), https://doi.org/10.1103/PhysRevB.76.125419

\bibitem{penn} David R. Penn, Phys. Rev. \textbf{142}, 350 (1966), https://doi.org/10.1103/PhysRev.142.350

\bibitem{izyum} Yu. A. Izyumov, Sov. Phys. Usp. \textbf{38}, 385 (1995), http://dx.doi.org/10.1070/PU1995v038n04ABEH000081




\end{thebibliography}
\end{document}